\documentclass[a4paper]{aa}
\usepackage{psfig}
\usepackage{amssymb}
\psfigurepath{fig}
\def\amin{\ifmmode ^{\prime}\else$^{\prime}$\fi}
\def\asec{\ifmmode ^{\prime\prime}\else$^{\prime\prime}$\fi}

\def\etal{{\it et\,al.\,}}
\def\mic{$\,\mu\rm m$~}
\renewcommand{\theenumi}{\roman{enumi}}

\begin{document}
\thesaurus{11.03.4;   
           11.05.2;   
           12.03.3;   
           13.09.1    
           }
\title{
An excess of mid-IR luminous galaxies in Abell 1689 ?
\thanks{Based on observations with ISO, an ESA project with instruments
funded by ESA Member States (especially the PI countries: France,
Germany, the Netherlands and the United Kingdom) with the participation
of ISAS and NASA}
}

\author{D. Fadda\inst{1} \and D. Elbaz\inst{1} \and P.--A. Duc\inst{2,3}
\and H. Flores\inst{1} \and A. Franceschini\inst{4} \and
C. J. Cesarsky\inst{5} \and A. F. M. Moorwood\inst{5} } 
\institute{ CEA Saclay - Service
d'Astrophysique, Orme des Merisiers, 91191 Gif-sur-Yvette C\'edex, France \and
CNRS URA 2052 and CEA Saclay - Service
d'Astrophysique, Orme des Merisiers, 91191 Gif-sur-Yvette C\'edex, France \and
University of Cambridge, Institute of Astronomy, Madingley Road, 
Cambridge, CB3~0HA, UK \and
Dipartimento di Astronomia, Universit\`a di Padova, Vicolo
dell'Osservatorio, 5, I35122 Padova, Italy \and 
European Southern Observatory, Karl-Schwarzschild-Strasse 2,
 D-85748 Garching bei M\"unchen, Germany }

\offprints{dfadda@cea.fr}
  
\date{\today}

\titlerunning{MIR emission from A1689}
\authorrunning{D. Fadda et al.}    

\maketitle

\begin{abstract}  
  
  We present the results of  infrared observations of Abell 1689,
  an exceptionally rich cluster of galaxies at intermediate redshift
  ($z\simeq 0.181$ ). It was observed with ISOCAM, at 6.7\mic and
  15\mic, and ISOPHOT at 200\mic from the {\it Infrared Space
    Observatory} (ISO). The cluster galaxies detected above a
  sensitivity limit of 0.15 mJy in the 6.7\mic band, whose emission is
  mostly dominated by their stellar component, show optical colors
  similar to the overall cluster population and are gathered in the
  center of the cluster, following the distribution of the cluster
  early-types.  In the 15\mic band, above a sensitivity limit of 0.3
  mJy, the galaxies spectroscopically confirmed to be cluster members
  are blue outliers of the cluster color-magnitude relation and become
  brighter going from the center to the outer parts of the cluster.
  
  We compare the 6.7\mic and 15\mic fluxes and the cumulative
    distributions of the B-[6.75] and B-[15] colors of the A1689
  galaxies, above our 90\% completeness limits of 0.2 and 0.4 mJy for
  6.7\mic and 15\mic respectively, to the galaxies of two nearby
  clusters, Virgo and Coma, and to the field galaxies at the same
  redshift of the cluster.   Although the B-[6.7] color
    distributions of the three clusters are compatible, we find a systematic
    excess of B-[15] color distribution for the galaxies located in
  Abell 1689 with respect to Coma or Virgo galaxies. This result proves the
  existence of a mid-infrared equivalent of the Butcher-Oemler effect
  measured in the optical. The comparison of 15\mic flux and
    B-[15] color distributions of A1689 and field galaxies does not
  show strong differences between the population of starburst 
  galaxies in the cluster and in the field.

\end{abstract}

\section{Introduction}
The first evidence for an increase of the star formation activity in
galaxies within high redshift clusters with respect to nearby clusters
was the discovery of an increasing fraction of blue galaxies with
cluster redshift ( Butcher \& Oemler 1978, 1984).  This effect was
later confirmed by other photometric studies (e.g. Couch \& Newell
1984, Rakos \& Schombert 1995, Lubin 1996, Margoniner \& de Carvalho
2000) and searches for spectroscopic signatures of the star formation
activity (Couch \& Sharples 1987, Dressler, Gunn and Schneider 1985,
Dressler \& Gunn 1992).

All these studies are limited to optical data and thus they have no
access to the global star formation activity. Only infrared  or
  radio data (see e.g. Smail et al. 1999)  can in fact reveal the
part of star formation activity hidden by dust.  But up to now, the
Butcher--Oemler effect has never been studied in the mid-infrared.
The high sensitivity and resolution of ISOCAM (Cesarsky \etal 1996a)
on board of the ISO satellite (Kessler \etal 1996) has allowed the MIR
observation of galaxies in distant clusters for the first time.  The
analysis of these new data will shed new light on this issue and
constrain the model predictions.

The usual scenario to explain the Butcher-Oemler effect invokes the
interaction of infalling galaxies with the dense cluster environment.
However, the specific role of the environment on the activity of these
galaxies has been subject to debate since it could either enhance star
formation or, on the contrary, quench it. Indeed, star formation could
be triggered by galaxy-galaxy interaction (e.g. Lavery \& Henry 1986),
galaxy harassment (Moore \etal 1996) or interaction with the
intracluster medium (ICM, see Gunn \& Gott 1972; Gavazzi \& Jaffe
1987), but it could also be quenched by ram pressure or tidal
stripping (e.g. Ghigna \etal 1998, Ramirez \& de Souza 1998, Fujita \&
Nagashima 1999).

Comparing a large sample of galaxies in medium-redshifts clusters to
field galaxies, Balogh \etal (1998) found that galaxies with similar
bulge-to-total luminosity ratio show stronger $[OII]\lambda 3727$ equivalent
widths, hence higher star formation rates, in the field than in
clusters. They conclude that the dense environment of cluster galaxies
quenches their star formation activity, independently on the 
morphological type of the galaxies. Abraham \etal (1996) reached the
same conclusion in the study of a large strip of sky centered on the
rich cluster Abell 2390 ($z=0.23$).

Dressler \etal (1999) observed another sample of ten distant rich
galaxy clusters. Using HST images to deduce the galaxy morphologies,
they definitively proved that cluster galaxies have reduced star formation
 with respect to field galaxies with the same morphological
types.

Galaxies with no strong dust extinction show a correlation of their
MIR flux densities with UV (Boselli \etal 1997) or $H\alpha$ (Roussel
\etal 2000).  However, the strongest starbursts known do not present
such correlation because they are heavily extincted (Rigopoulou \etal
1999) and they emit most of their light in the infrared (Sanders \&
Mirabel 1996).  Poggianti \& Wu (1999) have shown that almost 50\% of
very luminous IR galaxies present spectra with both strong
$H_{\delta}$ in absorption and relative modest [OII] emission.  These
galaxies, thought to be dusty starburst, are detected also in distant
clusters (Poggianti \etal 1999) and are expected to be easily
detectable by ISOCAM in the clusters as well as in the case of deep fields
(see Flores \etal 1999).

To explore the role of the cluster environment in galaxy evolution we
have therefore performed a survey of clusters from the nearby universe
to $z\sim 1$ (Fadda \& Elbaz 1998). In this paper, we present the
results of our MIR and FIR observations of Abell 1689, an
exceptionally rich cluster of galaxies at z=0.181 (Struble \& Rood
1987), with respectively ISOCAM and ISOPHOT
(Lemke \etal 1996) on-board ISO. 

Although the evolution of the fraction of blue galaxies at low redshifts
($0.1<z<0.2$) is not yet well established (see Butcher \& Oemler
1984, Rakos \& Schombert 1995 and Margoniner \& de Carvalho 2000),
A1689 shows a clear Butcher-Oemler effect since its fraction of blue
galaxies is about three times higher than in nearby clusters (Butcher
\& Oemler 1984).  This value, which has been questioned by the study
of Gudehus \& Hegyi (1991), has been recently confirmed by
Margoniner \& de Carvalho (2000).

We compare in the present paper the star formation activity, based on MIR
fluxes, of the Abell 1689 galaxies with that in the
nearby rich clusters  Coma and Virgo. 

After the presentation of the MIR observations and data reduction,
we describe the IR properties of the A1689 galaxies through their
radial and color distribution, we compare their MIR
luminosity functions with those of the Coma and Virgo clusters
and of field galaxies.  In a companion paper (Duc \etal 2000), we will
present the optical follow-up observations and an analysis of the spectral
properties, morphology and star formation of the galaxies detected.
Throughout all the paper, we assume $q_0 = 0.5$ and $H_0 = 75 \ km \ 
s^{-1} \ Mpc^{-1}$.
\begin{table}[t!]
\setlength{\tabcolsep}{1.8mm}
\begin{tabular}{llllccl}
\hline
Filter & Field & N$_{f}$ & N$_{r}$ & T$_{int}$ & PFoV & Obs. Nr.\\
\hline
CAM-LW3$^{*}$&6\amin  x6\amin& 9& 419 & 5s & 6\asec& 06600703\\
CAM-LW3$^{*}$&6\amin  x6\amin& 9& 308 &10s & 6\asec& 06600602\\
CAM-LW2$^{*}$&6\amin  x6\amin& 9& 292 &10s & 6\asec& 07000407\\
CAM-LW3      &5\amin  x5\amin&16& 462 &10s & 6\asec& 25402507\\
CAM-LW2      &5\amin  x5\amin&16& 462 &10s & 6\asec& 25402507\\
PHT-C200     &9\amin x12\amin& 6&  16 &32s &90\asec& 40100233\\
\hline
\end{tabular}
\caption{ISOCAM Observations of A1689 presented in the paper.
Asterisks mark observations executed during the performance verification phase.
We report the  area observed, the number of raster pointings or frames ($N_{f}$), 
the number of readouts per pointing ($N_{r}$) or ramps in the case of C200,
the integration time, the pixel field of view and the observation number in the ISO archive.
}
\label{tab-iso}
\end{table}

\section{Observations and Data Reduction}
\subsection{Observations}

\begin{figure*}[t!]
\vspace*{17.5cm}
\caption{Finding chart of the sources detected by ISOCAM. The image was taken
  at the NTT in the R band (see Duc \etal 2000).  Galaxies identified
  by their numbers are listed in Table 2. Bold slanted numbers
  indicate confirmed cluster members. Units for the axes are
  arcminutes.}
\label{ima:finding_chart}
\end{figure*}

Abell 1689 has been observed with ISO as part of a program of
observations of distant galaxy clusters (DEEPZCLS, P.I.: A.
Franceschini).  The observations were performed in the two MIR bands
LW2 and LW3 which cover the spectral regions 5.0--8.3\mic and
11.6-18.0\mic, respectively and are centered at 6.75\mic and 15\mic.
Additionally, the cluster was observed in the broad FIR band C200
which covers the spectral region 187--217\mic and is centered at
200\mic.  Table~\ref{tab-iso} summarizes the main features of these
observations.  

\begin{table*}[ht!]
\caption{List of Sources}
\scriptsize
\setlength{\tabcolsep}{1.6mm}
\begin{tabular}{rcc ccll  ccrr ccrr}
\hline
\\
\multicolumn{6}{c}{Optical counterpart}&
\multicolumn{1}{c}{Mem-}&
\multicolumn{4}{c}{LW2}&
\multicolumn{4}{c}{LW3}
\\
\multicolumn{1}{c}{Id.}&
\multicolumn{1}{c}{RA}&
\multicolumn{1}{c}{DEC}&
\multicolumn{1}{c}{B}&
\multicolumn{1}{c}{R}&
\multicolumn{1}{c}{z}&
\multicolumn{1}{c}{ber-}&
\multicolumn{1}{c}{offset}&
\multicolumn{1}{c}{flux}&
\multicolumn{1}{c}{SNR}&
\multicolumn{1}{c}{p}&
\multicolumn{1}{c}{offset}&
\multicolumn{1}{c}{flux}&
\multicolumn{1}{c}{SNR}&
\multicolumn{1}{c}{p}
\\
\multicolumn{1}{c}{Nr.}&
\multicolumn{1}{c}{(2000)}&
\multicolumn{1}{c}{(2000)}&
\multicolumn{1}{c}{}&
\multicolumn{1}{c}{}&
\multicolumn{1}{c}{}&
\multicolumn{1}{c}{ship}&
\multicolumn{1}{c}{(arcsec)}&
\multicolumn{1}{c}{($\mu$Jy)}&
\multicolumn{1}{c}{}&
\multicolumn{1}{c}{}&
\multicolumn{1}{c}{(arcsec)}&
\multicolumn{1}{c}{($\mu$Jy)}&
\multicolumn{1}{c}{}&
\multicolumn{1}{c}{}
\\
\multicolumn{1}{c}{(1)}&
\multicolumn{1}{c}{(2)}&
\multicolumn{1}{c}{(3)}&
\multicolumn{1}{c}{(4)}&
\multicolumn{1}{c}{(5)}&
\multicolumn{1}{c}{(6)}&
\multicolumn{1}{c}{(7)}&
\multicolumn{1}{c}{(8)}&
\multicolumn{1}{c}{(9)}&
\multicolumn{1}{c}{(10)}&
\multicolumn{1}{c}{(11)}&
\multicolumn{1}{c}{(12)}&
\multicolumn{1}{c}{(13)}&
\multicolumn{1}{c}{(14)}&
\multicolumn{1}{c}{(15)}
\\
\\
\hline
\\
1 &$\ \ $$\ \ $$\ \ $$\ \ $&$\ \ $$\ \ $$\ \ $$\ \ $&18.8&18.0&2.584 &B&$+0.0 -0.4$&$ 870\pm120$&16&$<$.001&$-1.1 +2.1$&$2600\pm360$&16&$<$.001\\
2 &&&18.2&15.8&0     &S&$+2.4 -1.2$&$1300\pm140$&20&$<$.001&$         $&$<770$      &  &     \\
3 &&&19.8&17.4&0.192 &M&$-0.3 +2.6$&$ 313\pm 70$& 8&   .001&$+3.0 +1.7$&$ 450\pm180$& 5&   .001\\
4 &&&20.8&19.1&0.215 &M&$+0.6 +1.3$&$<120$      & 2&   .003&$+1.3 -1.1$&$ 320\pm230$& 4&$<$.001\\
5 &&&19.4&17.6&0.086 &F&$+1.6 +1.6$&$ 190\pm 70$& 7&   .001&$-2.0 -1.1$&$ 440\pm180$& 5&$<$.001\\
6 &&&20.6&19.1&0.216 &M&$-3.3 -1.3$&$ 190\pm 70$& 5&   .001&$+1.4 -1.0$&$<280$      & 2&   .004\\
7 &&&20.3&17.8&0.19$^{ph}$ &M&$-0.8 +2.4$&$ 190\pm 70$& 6&   .001&$         $&$<280$      &  & \\
8 &&&20.3&18.0&0.171 &M&$+2.2 -1.5$&$ 260\pm 70$& 7&   .003&$         $&$<280$      &  &       \\
9 &&&14.6&$<$14&0    &S&$+0.9 +0.5$&$1770\pm180$&48&$<$.001&$         $&$<360$      &  &       \\
10&&&20.3&18.3&0.178 &M&$+1.0 -1.0$&$ 200\pm 70$& 6&$<$.001&$+0.8 -0.9$&$ 600\pm190$& 7&$<$.001\\
11&&&21.1&18.5&0.20$^{ph}$ &M&$+0.5 -0.3$&$ 200\pm 70$& 6&$<$.001&$         $&$<280$      &  & \\
12&&&20.3&18.8&0.173 &M&$-1.4 +0.4$&$ 190\pm 70$& 3&$<$.001&$         $&$<360$      &  &       \\
13&&&21.0&18.6&0.18$^{ph}$ &M&$       ^*$&$ 130\pm 70$& 4&   $^*$&$         $&$<280$      &  & \\
14&&&18.4&16.0&0.184 &M&$       ^*$&$ 490\pm 90$&17&   $^*$&$-2.9 +2.1$&$ 460\pm180$& 5&$<$.001\\
15&&&21.0&19.0&0.397 &B&$+2.1 +0.4$&$ 190\pm 70$& 3&$<$.001&$-0.2 +0.5$&$ 900\pm170$& 6&$<$.001\\
16&&&19.8&18.1&0.200 &M&$       ^*$&$ 300\pm 70$& 9&   $^*$&$         $&$<280$      &  &       \\
17&&&20.4&17.8&0.19$^{ph}$ &M&$       ^*$&$ 300\pm 70$& 9&   $^*$&$-0.2 -2.4$&$ 430\pm190$& 5&   .001\\
18&&&18.9&17.3&0.204 &M&$       ^*$&$ 230\pm 70$& 7&   $^*$&$         $&$ <280$     &  &       \\
19&&&20.4&17.7&0.19$^{ph}$  &M&$       ^*$&$ 170\pm 70$& 5&   $^*$&$         $&$ <280$     &  &\\
20&&&19.5&17.2&0.174 &M&$       ^*$&$ 400\pm 80$&13&   $^*$&$         $&$ <280$     &  &       \\
21&&&20.9&18.2&0.19$^{ph}$  &M&$       ^*$&$ 220\pm 70$& 6&   $^*$&$-1.5 -1.5$&$ 330\pm230$& 4&   .002\\
22&&&20.4&17.9&0.192 &M&$       ^*$&$ 100\pm 60$& 4&   $^*$&$         $&$ <280$     &  &       \\
23&&&21.9&20.6&0.692 &B&$         $&$<100$      &  &       &$+3.2 +0.3$&$ 340\pm220$& 4&   .020\\
24&&&20.9&18.3&0.197 &M&$-1.9 +0.9$&$ 200\pm 70$& 6&   .001&$         $&$ <280$     &  &       \\
25&&&15.9&15.3&0     &S&$+0.9 +1.5$&$ 290\pm 70$& 9&$<$.001&$         $&$ <280$     &  &       \\
26&&&19.5&17.0&0.188 &M&$+0.4 +0.4$&$ 330\pm 70$&11&$<$.001&$-0.6 +0.6$&$ <280$     & 3&$<$.001\\
27&&& -  &21.6&0.87$^{ph}$ &B&$         $&$<110$      &  &       &$-0.6 -0.8$&$ 820\pm170$& 8&   .005\\
28&&&20.2&18.7&0.15$^{ph}$ &M?&$         $&$<90$       &  &       &$-0.5 +0.4$&$ 400\pm200$& 5&   .001\\
29&&&20.6&18.3&0.177 &M&$+2.1 +0.6$&$ 100\pm 60$& 3&$<$.001&$         $&$ <280$     &  &       \\
30&&&20.4&17.7&0.180 &M&$+0.0 +2.1$&$ 210\pm 70$& 7&$<$.001&$-3.4 -1.2$&$ <280$     & 3&   .001\\
31&&&19.4&17.6&0.176 &M&$+0.0 +1.0$&$ 110\pm 60$& 3&$<$.001&$+0.4 +0.7$&$ 600\pm190$& 6&$<$.001\\
32&&&19.1&16.6&0.202 &M&$+1.0 +0.8$&$ 350\pm 70$&13&$<$.001&$+0.2 +1.8$&$ <280$     & 3&$<$.001\\
33&&&19.5&16.8&0.201 &M&$+0.2 -0.3$&$ 230\pm 70$& 7&$<$.001&$         $&$ <280$     &  &       \\
34&&&17.2&16.0&0     &S&$-0.4 +2.1$&$ 320\pm 70$&15&$<$.001&$-1.7 -1.1$&$ <280$     & 2&       \\
35&&&19.7&17.9&0.181 &M&$-1.5 -0.4$&$ 210\pm 70$& 6&   .001&$-0.9 -0.3$&$<400$      & 3&   .001\\
36&&&17.3&15.9&0     &S&$-0.9 +1.0$&$ 240\pm 70$& 8&$<$.001&$         $&$<280$      &  &       \\
37&&&19.7&17.7&0.199 &M&$-2.3 +2.6$&$ 220\pm 70$& 4&   .002&$-1.0 +0.8$&$1970\pm300$&16&$<$.001\\
38&&&22.5:&19.1&-    &?&$         $&$<170$      &  &       &$-0.9 +2.9$&$ 680\pm190$& 5&   .006\\
39&&&22.6&20.4&0.19$^{ph}$ &M&$         $&$<130$      &  &       &$-1.2 -0.7$&$ 370\pm210$& 4&   .012\\
40&&&20.4&17.9&0.187 &M&$+0.6 +4.7$&$ 150\pm 70$& 4&   .003&$         $&$ <280$     &  &       \\
41&&&19.9&18.3&0.200 &M&$-1.2 +1.1$&$ 190\pm 70$& 5&$<$.001&$-0.2 +1.0$&$790\pm170$ & 8&$<$.001\\
42&&&21.1&19.3&0.13$^{ph}$ &F?&$+3.2 +0.5$&$ 110\pm 60$& 3&   .004&$+1.1 -2.7$&$<280$      & 3&   .009\\
43&&&20.2&18.2&0.196 &M&$-4.9 +1.9$&$ 160\pm 70$& 4&   .008&$         $&$<280$      &  &       \\
44&&&17.3&15.7&0     &S&$-2.0 +0.5$&$ 450\pm 80$&11&$<$.001&$         $&$<320$      &  &       \\
45&&&19.9&18.7&0.188 &M&$+0.1 +2.6$&$ 490\pm 90$&10&   .002&$+1.4 +2.1$&$1230\pm220$& 7&   .001\\
\\
\hline\\                                                                                                            
\end{tabular}\\                                                                                                    
\noindent{\footnotesize {\em Comments}: For each detection we list:
the identification number (col.~1), the coordinates of the optical counterpart (cols.~2 and~3), the B and R magnitudes (cols.~4 and~5), the redshift (col.~6) and the membership (col.~7: M = member, F = foreground, B = background, S = star). 
The B filter corresponds  to the $B_b$ filter of the EMMI camera on the ESO-NTT telescope.
For the two ISOCAM filters we report the offset (cols.~8 and~12) from the optical counterpart, the flux with the error (or the
3$\sigma$ upper limit, cols.~9 and~13), the signal to noise ratio (cols.~10 and~14) and the probability of spuriuos identification, p (cols.~11 and~15). 
Asterisks in col.~11 mark LW2 sources in the central concentration which have been deconvolved.
While the SNR is computed on the image, the flux error is dominated by
the uncertainty in the correction factor computed through simulations (see Figure~8 and the text).  All the redshifts are from Duc et al. (2000), except for \#1, which is a QSO, from Hewett, Foltz \& Chaffee (1995), \#29, from  Teague, Carter \& Gray (1990) and photometric redshifts (marked with a $ph$) from Dye \etal (2000). 
}\\
\label{tab:list}
\end{table*}


The field has been covered with subsequent pointings of the ISO camera
(raster).  For each pointing, $\sim 25$ readouts were performed since
the detector needs a time to stabilize which depends on the mean
flux and on the flux of the source. Since Abell 1689 is near to the
ecliptic plane ($\lambda = $ 197.00, $\beta = $ 5.78) where the
zodiacal light is intense, the mean flux is high and the signal always
reaches stabilization (Coulais \& Abergel 1999).

Some observations of A1689 were performed during the performance
verification phase of the ISO mission in order to determine the best
combination of parameters to observe faint sources with ISOCAM. We
have used only three of these observations, because the quality of the
other is very poor.

The total time spent for each pixel of the final image, which takes into
account all the rasters coadded, is about 25 minutes for the LW2 filter,
30 minutes for the LW3 filter and 77 seconds for the C200 filter.

\begin{figure*}[t!]
\vspace*{22cm}
\caption{Galaxies detected with the LW2 and LW3 filters out of the cluster center on HST images. For each galaxy, LW2 and LW3 contours are on the left and on
  the right, respectively.  Contour levels are expressed in $\mu$Jy
  arcsec$^{-2}$, while axes units are in arcseconds. Numbers refer to
  Table~2.}
\label{ima:sources_common}
\end{figure*}
\begin{figure*}[t!]
\hbox{
\vspace*{7cm}
}
\caption{HST image of the central crowded area of A1689 superposed with LW2 (left) and LW3 (right) contours.
  Contour levels are expressed in $\mu$Jy arcsec$^{-2}$ and axes units
  in arcsecons.  Galaxies identified by their numbers are listed in
  Table 2.}
\label{ima:center}
\end{figure*}
\begin{figure}[b!]
\psfig{file=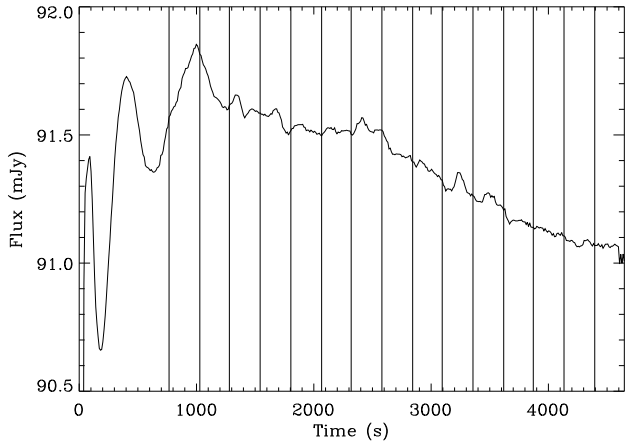,width=0.5\textwidth}
\caption{
The initial transient in the case of an LW3 observation of the cluster
as seen in the median behavior of all the camera pixels.
Vertical lines indicate the moment when the camera changes its pointing
during the raster. After an initial increase in the flux there are some
oscillations. The first 70 readouts (almost 800 seconds) are the so
called stabilization time which is rejected in the analysis. 
In this case, we reject other two pointings because of a further bump
in the signal. Then, the drift in the signal is easily corrected
by the PRETI software.
}
\label{ima:ini_trans}
\end{figure} 

\subsection{Data reduction}

The data reduction can be summarized in the following steps:\\
(a) dark subtraction, using the time-dependent dark correction in the
CIA\footnote{CIA is a joint development by the ESA Astrophysics
  Division and the ISOCAM Consortium. The ISOCAM Consortium is led by
  the ISOCAM PI, C. Cesarsky, Direction des Sciences de la Mati\`ere,
  C.E.A., France.}
package (Biviano \etal, 1999); \\
(b) deglitching of short glitches due to low energy cosmic rays,
using the Multi-resolution Median Transform technique (Starck \etal, 1999); \\
(c) removal of long glitch tails due to high energy cosmic rays,
mainly using the PRETI algorithm (Starck \etal, 1999);\\
(d) flat fielding and baseline subtraction to remove the long term
transients
in pixel signals,  using the PRETI algorithm;\\
(e) coaddition of mosaic images, taking into account the camera
distortion (Aussel \etal 1999), to obtain a final image with a pixel
size of 2 arcsec.

Some further improvements, described below, have been added to this
process in order to best exploit  this data set.

We have not considered some noisy data:
we have masked the border pixels which are badly
illuminated and we have not taken into account the first
readouts when the signal is not yet stabilized.  Indeed, as an
observation begins, the flux on the detector changes rapidly and
exhibits a transient behavior with an artifact which can mimic a
source, hence we mask the first readouts for each pixel (see
figure~\ref{ima:ini_trans}).
\begin{figure} [b!]
\psfig{file=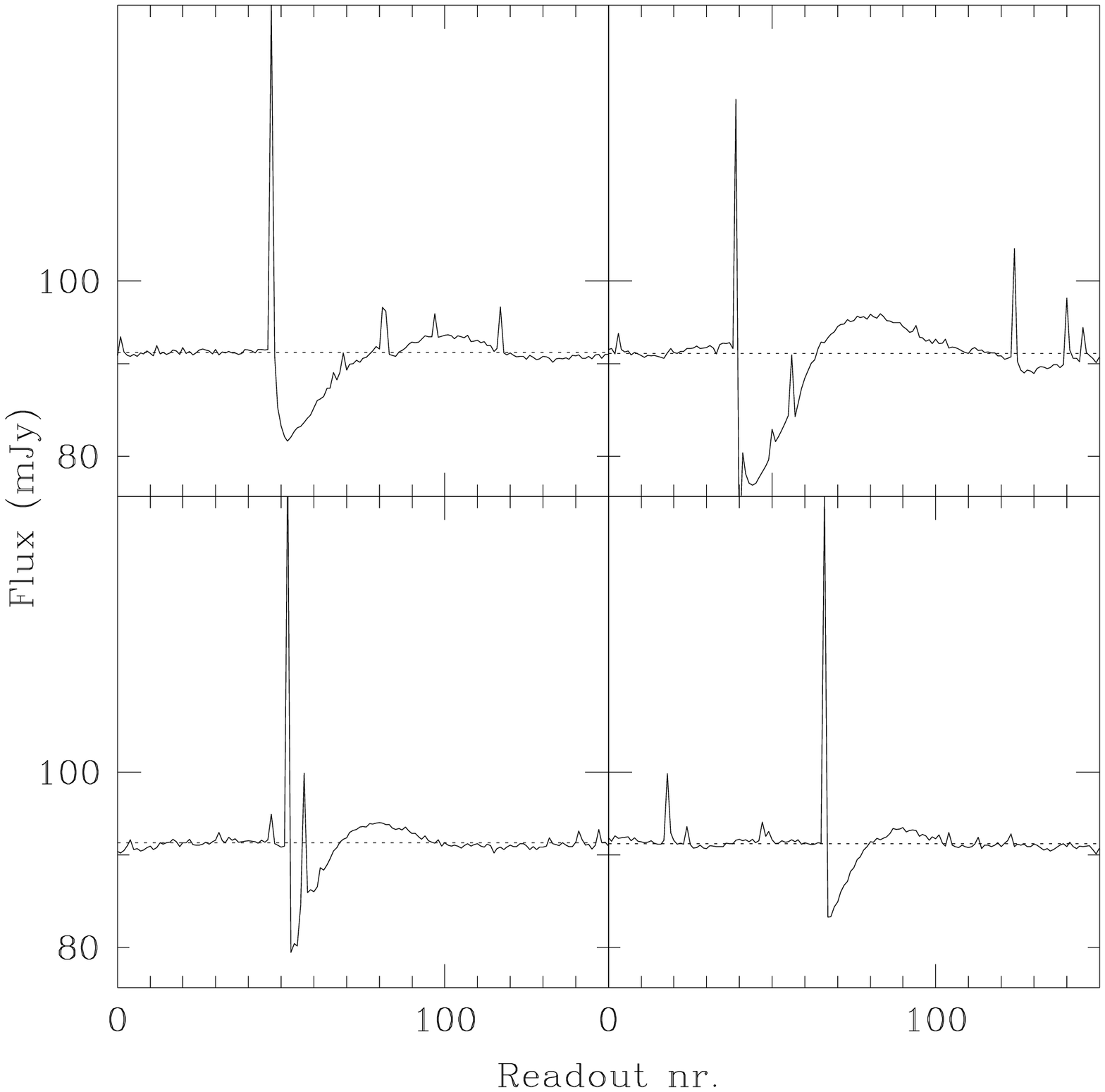,width=0.5\textwidth}
\caption{ Four examples of oscillations in the signal after an impact
of an energetic cosmic ray on the camera. After the glitch and its
negative tail the signal returns to the median value (dashed
line) passing through a positive bump that can be interpreted as a
source.}
\label{ima:bump_dipper}
\end{figure}

After a deep negative tail which follows the impact of an energetic
cosmic ray (see Starck et al., 1999), the detectors sometimes produce
a bump resembling a source. While it is easy for PRETI to remove the
negative oscillations, the first positive bump of the signal often has
the same temporal size as a typical source (see
figure~\ref{ima:bump_dipper}). Thus, when PRETI decomposes the signal
in components of different temporal scales, it classifies this bump as
a source introducing false detections. Unfortunately, due to the large
variety of cases and the presence of several glitches at the same
time, it is very difficult to correct the signal by modeling these type of glitches.
Therefore, we have decided to subtract from the signal each positive
component of the PRETI analyses which comes exactly after the negative
tail of a glitch.

\begin{figure*}[t!]
\hbox{
\vspace*{6.5cm}
}
\caption{HST images of galaxies detected only with the LW2 (left) or the 
  LW3 (right) filter.  Contour levels are expressed in $\mu$Jy
  arcsec$^{-2}$, while axes units are in arcseconds. Numbers refer to
  Table~2. Detections of \#40 and \#43 present large offsets because
  they are at the border of the field where redundancy is lower and
  distortion effects become important. The optical counterparts of
  \#23 and \#27, high redshift objects detected in the LW3 band, have
  distorted morphologies.  }
\label{ima:sources_sing}
\end{figure*}

Transient effects appear during and after observing a source because
of the inertia of the detector. The initial transient depends on the
mean flux and on the flux of the source and can affect the flux
measured.  In our case the mean flux is high because we are observing
an object near the ecliptic plan where the zodiacal light is intense
and the signal reaches stabilization after few readouts (Coulais \&
Abergel 1999). Therefore, we do not correct for this effect. Tails,
which appear after the detection of bright sources, can generate
artificial sources (ghosts) in the final image. Since transient
correction techniques (Coulais \& Abergel 1999) are not yet well
suited to correct this effect in the case of bright point sources, we
simply removed these tails from the signal.

Astrometry corrections must be made before coadding several
independent observations. Since the error in the angular position
of the camera is
negligible, the astrometry correction reduces to an offset. To
estimate the offsets we compare the ISO images with an optical image
and we evaluate the coordinate shifts of bright optical sources which
are detected in the field.  Offset corrections are 2 arcsecs on
average.

\section{Data analysis}
\subsection{Source Detection and Optical Counterparts}

We extract sources from the final image using the corresponding rms
map computed by PRETI (Starck \etal, 1999) which takes into account
the history of the signal for each pixel.  The use of the rms map in
this step is very important, because the sensitivity changes from
point to point due to the variable number of camera pointings and of
the number of the readouts which have been masked to remove glitches.

\begin{table} [!t]
\caption{Statistics of ISO detections.}
\begin{tabular}{llllll}
\hline
Filter & N$_{src}$ & N$_{z}$  & N$_{members}$ & N$_{bkg}$& N$_{stars}$\\
\hline
LW2     & 30 (39)& 30 (39)& 20(29) &4&6 \\
LW3     & 18     & 17& 12     &5&0 \\
LW2\&LW3& 12 (20)& 12 (20)&  9(14) &3(6)&(1) \\ 
\hline
\end{tabular}\\

\noindent{\footnotesize 
{\em Comments} : LW2 sources have been detected at the 3$\sigma$ level, LW3 sources at the 4$\sigma$ level. In the LW2 line,
the numbers in parentheses include the LW2 sources in the central region, which have been deblended. In the LW2\&LW3 line, the numbers in parentheses include also counterparts of LW2 or LW3 sources which have been detected at a lower $\sigma$ level than the previous ones.} 
\label{tab-ident}
\end{table}

To extract sources we consider the signal-to-noise ratio (SNR) map
computed using the ratio of the image and corresponding rms map and we
consider all the peaks greater than 3. We retain these sources if the
SNR computed for a region of 8 arcsec  radius is still greater than 3
for the LW2 sources. For LW3 we choose a more conservative level (SNR $>$ 4)
due to the higher frequency of glitches. 

After astrometry corrections, the error in global astrometry is
remarkably small: less than or equal to 2 arcsec. This allows us to
identify ISO sources with corresponding optical sources with a high
confidence level (see Figs.~\ref{ima:sources_common}, \ref{ima:center},
\ref{ima:sources_sing}).  To find optical
counterparts we have used a large field R-band image obtained with the
NTT at La Silla and images from the HST archive. For most of the
galaxies detected we have also obtained spectra with EMMI on the NTT
(Duc \etal, 2000). Optical counterparts are assigned with the method
described by Downes \etal (1986). For each source, we consider
candidates in the R-band image within a radius which depends on the
resolution of the observation (e.g. 6 arcseconds in the case of our
LW3 image).  Taking into account the fact that bright optical sources
are more easily detected than faint ones, the method assigns to each
candidate the probability of spurious identification.  This method
works well with point sources as in our case.

\begin{figure*}[t!]
\hbox{\psfig{file=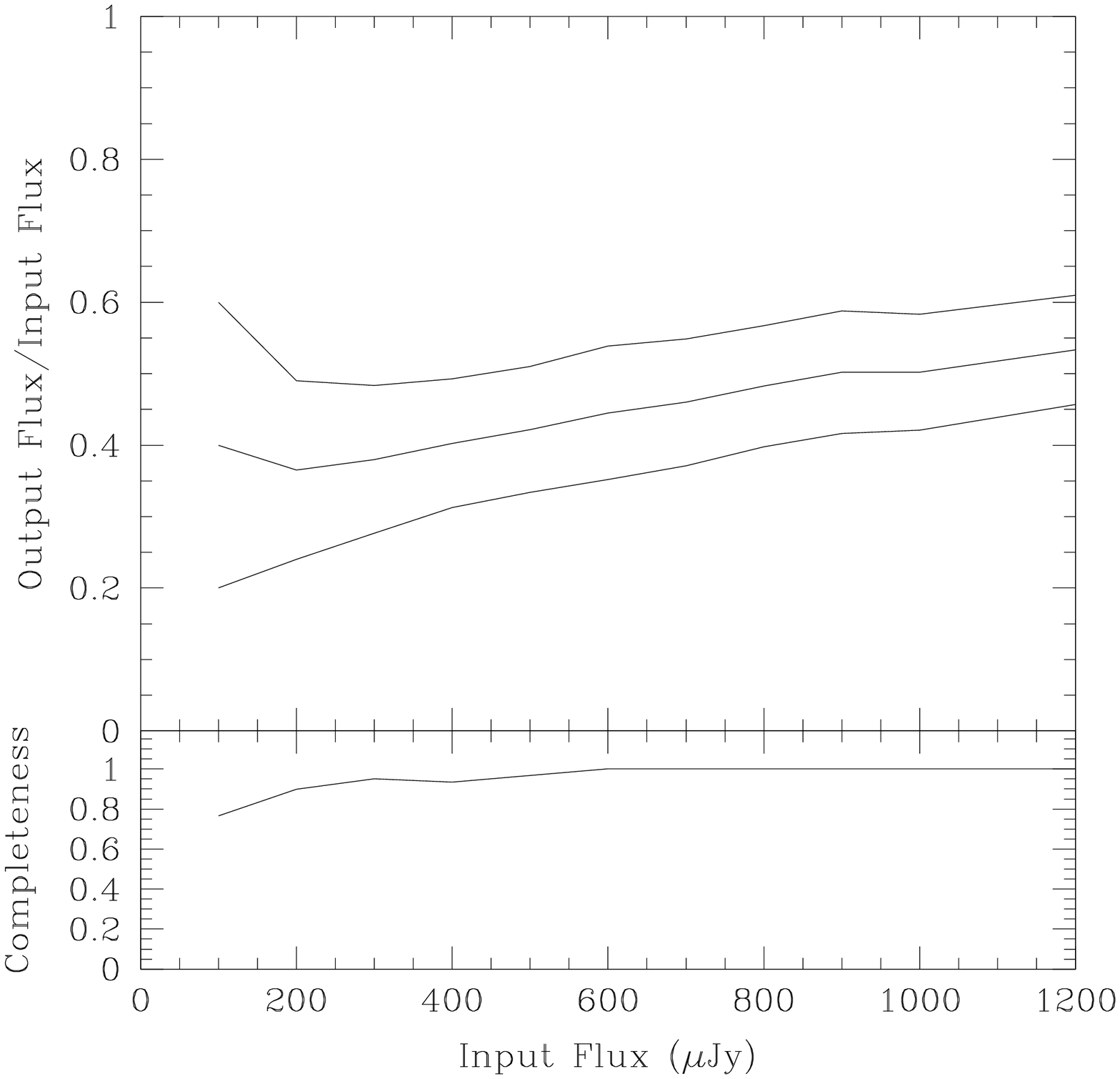,width=0.5\textwidth}
\psfig{file=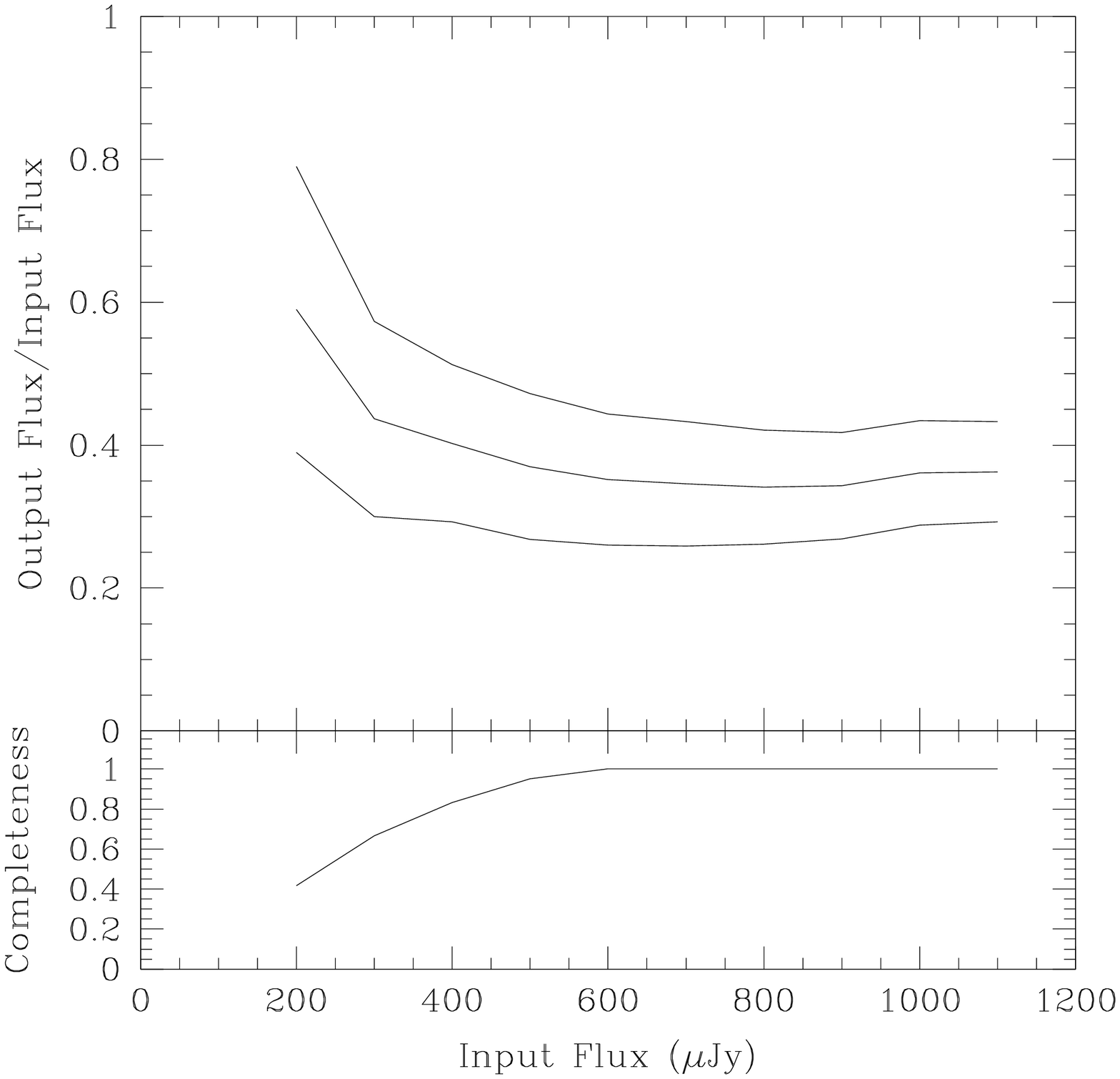,width=0.5\textwidth}}
\caption{
  Photometry and completeness results for LW2 (right) and LW3 (left)
  fluxes obtained through simulations. Top: the ratio between measured
  and input flux as a function of input flux with 1$\sigma$ error
  bands. Bottom: percentage of detected sources in simulations as a
  function of the input flux of the sources (at 3$\sigma$ level for
  LW2 and 4$\sigma$ level for LW3).  }
\label{ima:calibration}
\end{figure*}

For each detection, we report in Table~\ref{tab:list} the position of
the optical counterpart, its B and R magnitudes, the spectroscopic or
photometric redshift, the membership, the offsets from the optical
counterpart, the flux, the signal-to-noise ratio, and the probability
of spurious association with the optical source. If one source is
only detected in the LW2 (with SNR $>$ 3) or in the LW3 (with SNR $>$ 4)
filter we report the detection at lower SNR level in the other band
too. In case of no detection we report an upper limit (see below).

In the center of the LW2 image (see figure~\ref{ima:center}), several
sources are blended into one extended source. To estimate the flux of
the individual galaxies, we have assumed that the brightest optical
galaxies emit in the LW2 band. Then we have fitted ISOCAM point spread
functions at the positions of these sources by choosing the fluxes
which minimize the residual flux. LW2 fluxes obtained in such a way
are reported in Table~\ref{tab:list}.

Table~\ref{tab-ident}  summarizes the statistics of ISOCAM
detections.  We report, for each band, the number of detections with
optical counterparts, the number of sources with known redshift, the
spectroscopically confirmed number of cluster members and background
galaxies and the number of stars detected. 

A large number of detected galaxies are cluster members (88\% of LW2
sources and 66\% of LW3 sources), in contrast to that found by Altieri \etal
(1999) who found mainly lensed background galaxies for the core of A2390. 
The reason for this discrepancy is probably that deep images were obtained
by Altieri \etal for the very central part of A2390,
while the majority of the MIR cluster galaxies, harboring dust
obscured star formation, lie in the outskirts (see our results below).

\subsection{Photometry and Completeness}
Once the source positions are obtained, we performed aperture
photometry on the coadded image within a circle of radius 8 arcsec.
The fluxes obtained take into account only part of the PSF, because the
wings of the PSF are confused in the noise.  This effect depends on
the source flux and the wavelength. We have corrected the measured
fluxes through simulations.  

Simulations are made by introducing artificial
sources in a zone which is observed at least 25 minutes in the case of
LW2 observations and 30 minutes for the LW3 observations.  This region
covers most of the sources detected, except for few external high flux
sources. Choosing the positions we avoid regions with SNR $>$ 2.
We actually add these sources to the original data cube by taking into
account all the effects which enter in the real signal (dark current,
flat-field, transients, point spread functions and camera
distortions). We must use real data because a complete model of
noise including cosmic ray effects does not exist.  Then we analyze
this new cube exactly in the same manner as the real data and we
measure fluxes for sources which are detected.  In this way we are
able to produce a curve (see figure~\ref{ima:calibration}) which is
used to correct measured fluxes and assign an error or evaluate an
upper limit.
The uncertainty of this correction dominates the error in the flux 
evaluation on the image. In Table~\ref{tab:list} we report this error, while
the SNR is computed on the image. Therefore, also detections with high SNR
can have big photometric errors, especially in the case of low fluxes.

These simulations also allowed us to estimate the sensitivity limit and
the completeness for the two filters. In the LW2 band, it is possible
to detect sources to a sensitivity limit of 0.15 mJy while the
100\% completeness level is reached at 0.4 mJy.  The LW3 image, which
is noiser than the LW2 image due to higher zodiacal light contribution, has
a sensitivity limit of 0.3 mJy and is 100\% complete above 0.6 mJy.

\section{Results}
\begin{figure} [t!]
\psfig{file=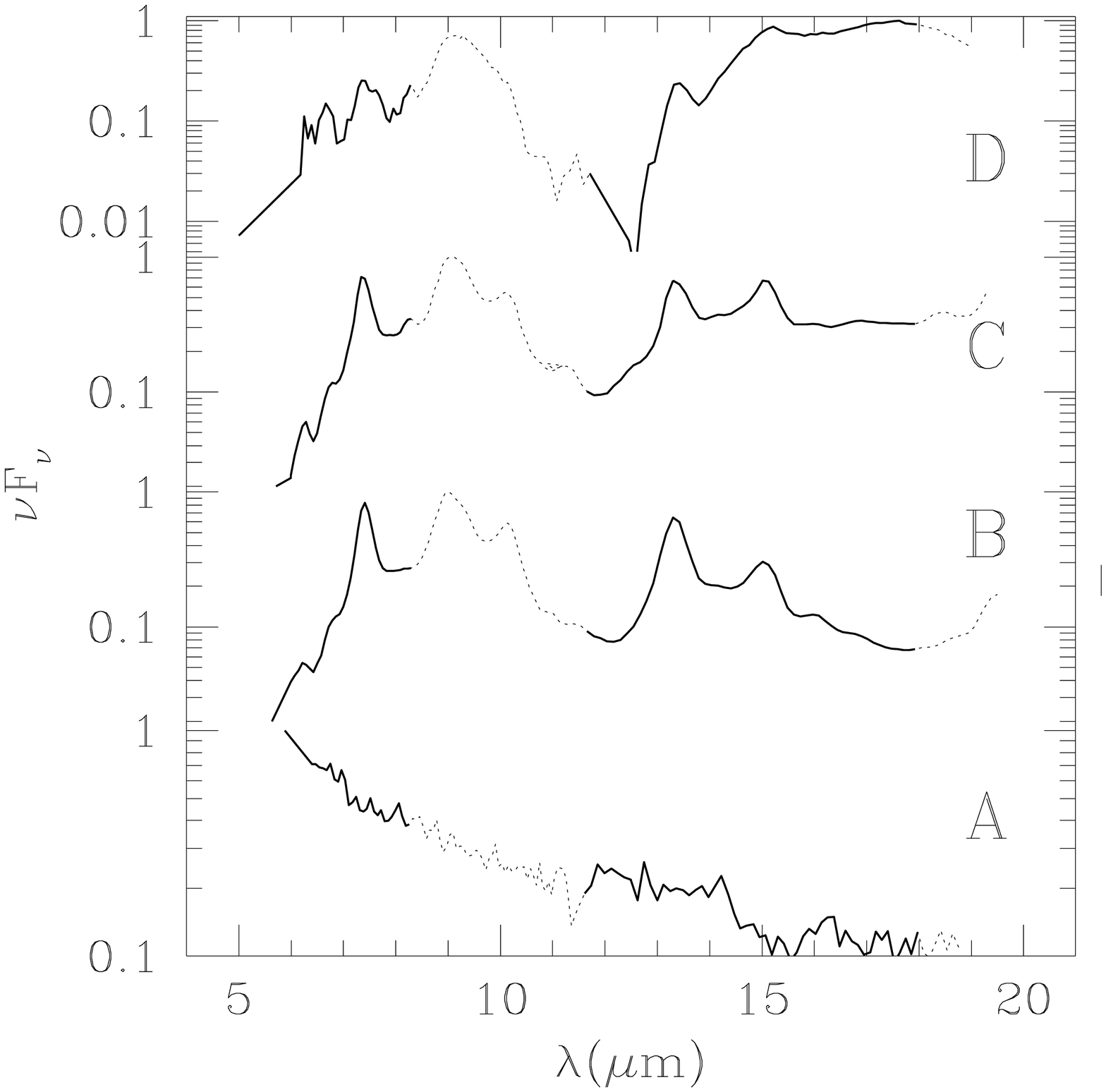,width=0.5\textwidth}
\caption{ MIR SEDs of: A) the elliptical galaxy NGC 1399 (Madden \etal, 1998),
B) the galactic photodissociation region  NGC 7023 (Cesarsky  \etal, 1996b), C) the starburst M 82
(Tran, 1998) and D) the ULIRG Arp 220 (Charmandaris \etal, 2000) placed
at the redshift of A1689 ($z=0.181$). $\nu F_{\nu}$ is in arbitrary
units.  Bold lines correspond to the spectral ranges seen by the two ISOCAM
filters LW2 (on the left) and LW3 (on the right).}
\label{ima:sed}
\end{figure}

The ISOCAM filters, LW2 and LW3, were designed to sample
two dominant components of the rest-frame MIR emission: broad bands
whose carriers are aromatic carbon compounds (hereafter called
aromatic features) and a continuum radiated by hot ($\sim 100 K$)
small dust grains stochastically heated (Cesarsky  \etal 1996a).
Template spectral emission distributions (SED) observed by the ISOCAM
CVF for different types of galaxies are shown in Figure~\ref{ima:sed}.
The spectral ranges of the SEDs observed by the LW2 and LW3 filters
are shown by bold lines.  The evolved stellar population dominates the
emission of a typical elliptical galaxy which does not exhibit MIR
features. Late-type galaxies show the aromatic features and the dust
continuum whose relative contribution increases with star formation
activity.  At the redshift of A1689, the LW2 filter can cover 
the tail of stellar emission and the 6.7\mic aromatic feature.
The LW3 band covers the 11.3\mic and 12.6\mic aromatic features and the hot 
dust continuum, which can dominate in galaxies  with a high
star formation rate.

\subsection{Radial Distribution}
 A rich, relaxed and unperturbed cluster has a radially decreasing density
profile of galaxies.  The distribution of galaxies with peculiar
properties can exhibit a spatial segregation, e.g.  morphological
(e.g. Dressler 1980) or a luminosity segregation (e.g.
Biviano \etal 1992).  In particular, starburst galaxies tend to be
more frequent in the outskirts of clusters as well as the blue
galaxies which are responsible for the Butcher-Oemler effect (Butcher
\& Oemler 1984, Abraham \etal 1996, Rakos, Odell \& Schombert, 1997,
Ellingson \etal 1999).  At the same time, emission line galaxies and blue
galaxies have  higher velocity dispersions and steeper velocity
dispersion profiles than quiescent and red galaxies, respectively
(Biviano et al. 1997, Carlberg et al. 1997), suggesting that
star-forming galaxies are falling into the cluster, maybe for the
first time.

\begin{figure}[t!]
\vbox{
\hbox{\psfig{file=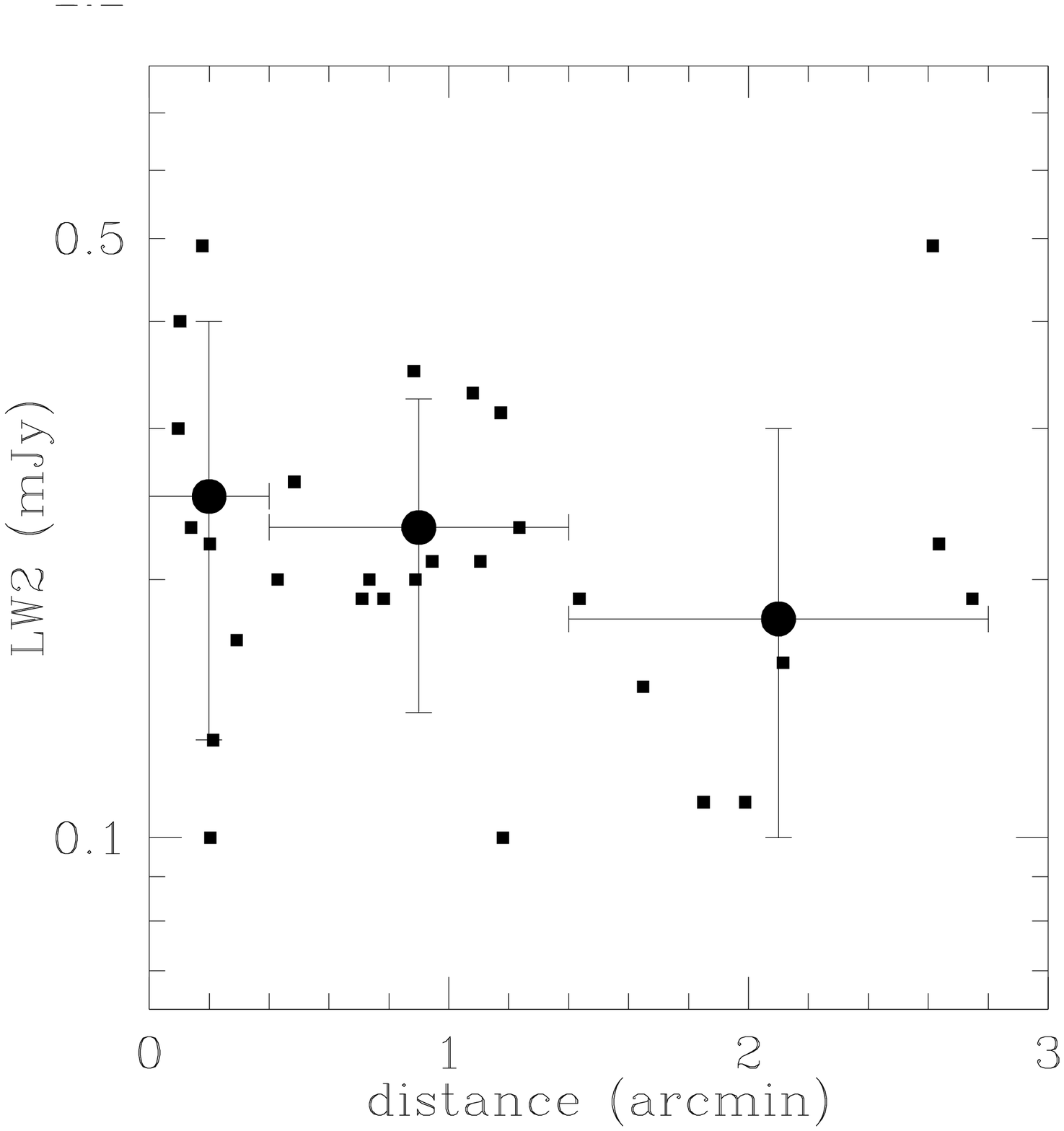,width=0.25\textwidth}
\psfig{file=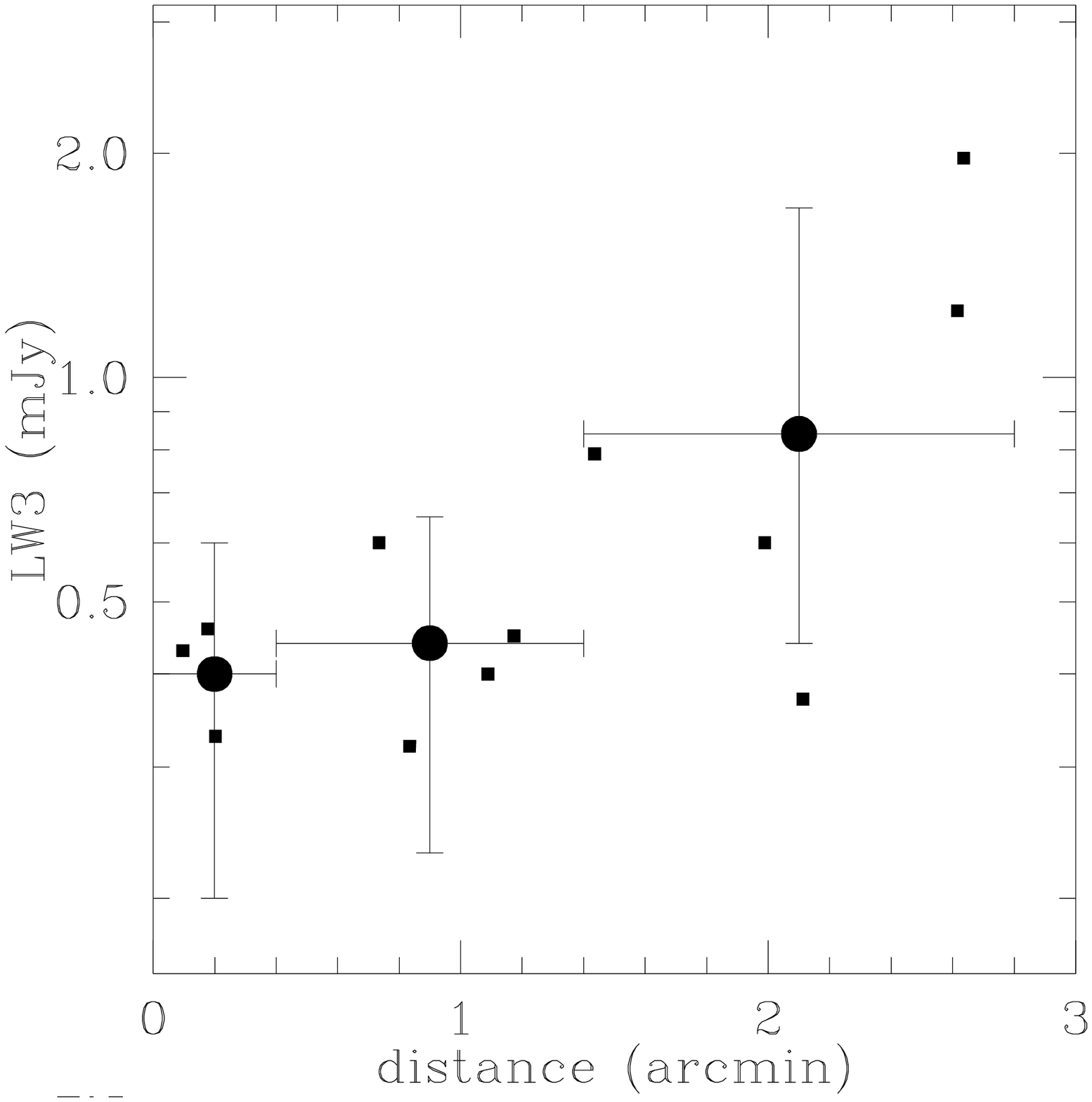,width=0.25\textwidth}}
\hbox{\psfig{file=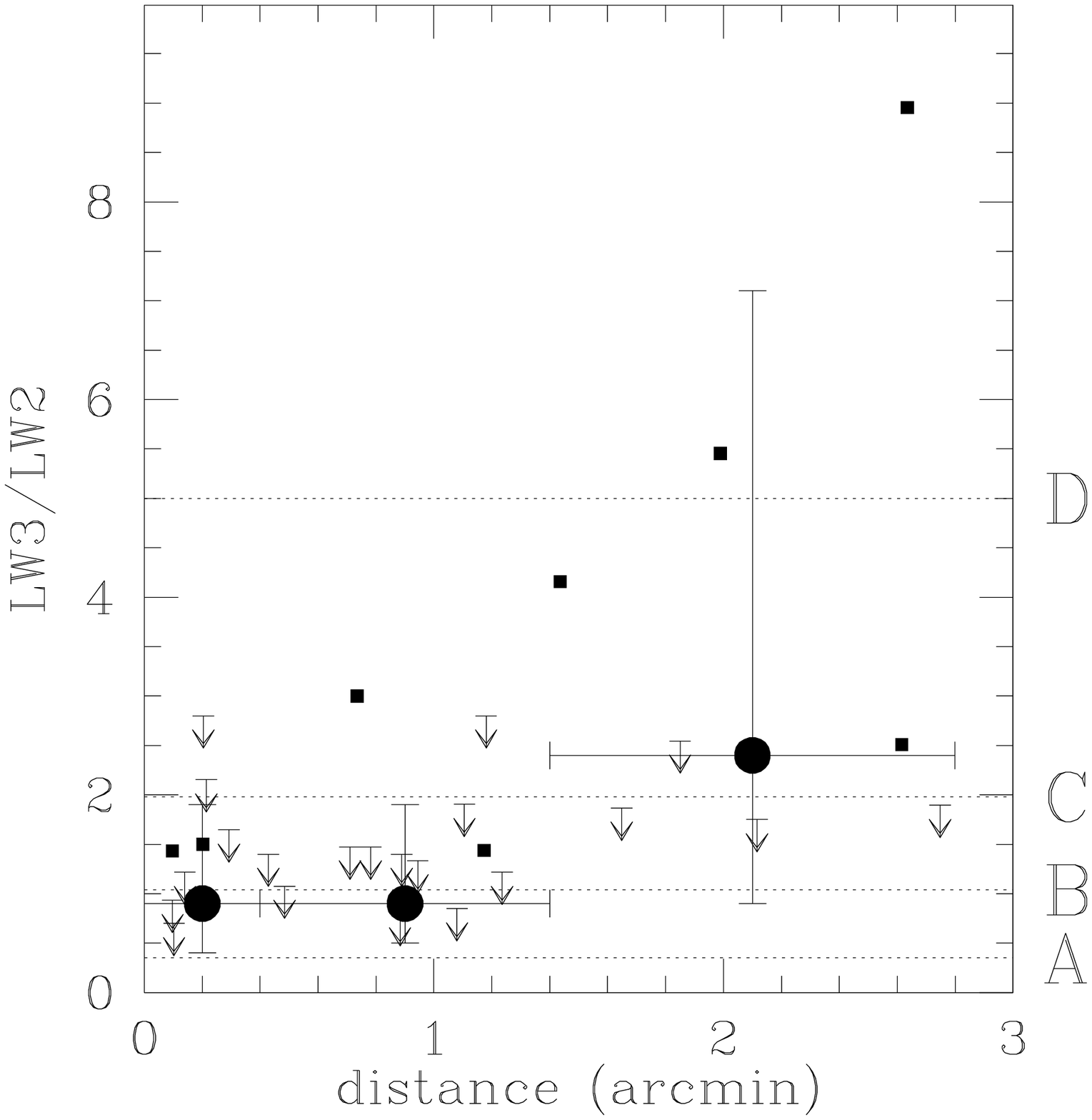,width=0.25\textwidth}
\psfig{file=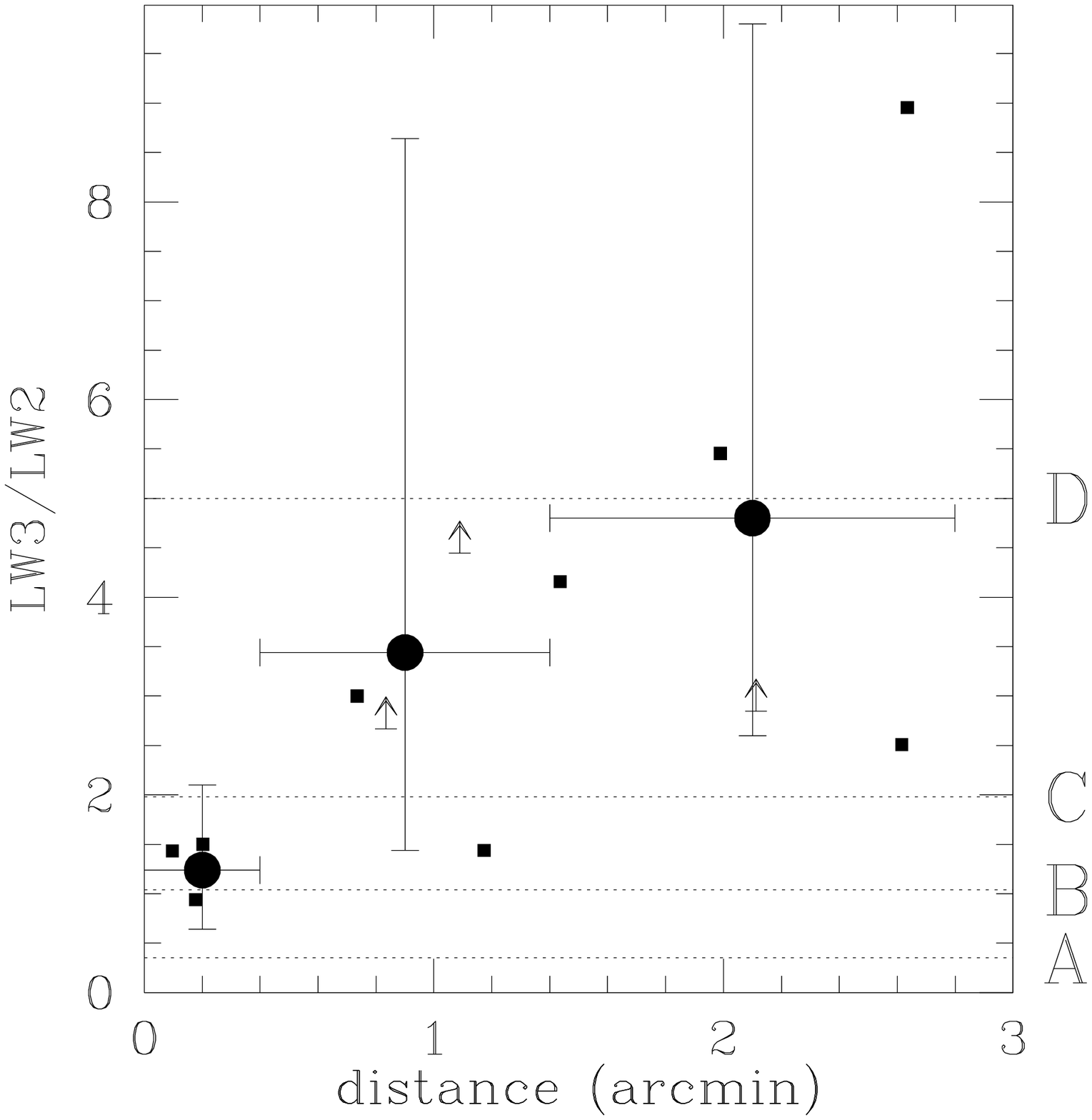,width=0.25\textwidth}}
\hbox{\psfig{file=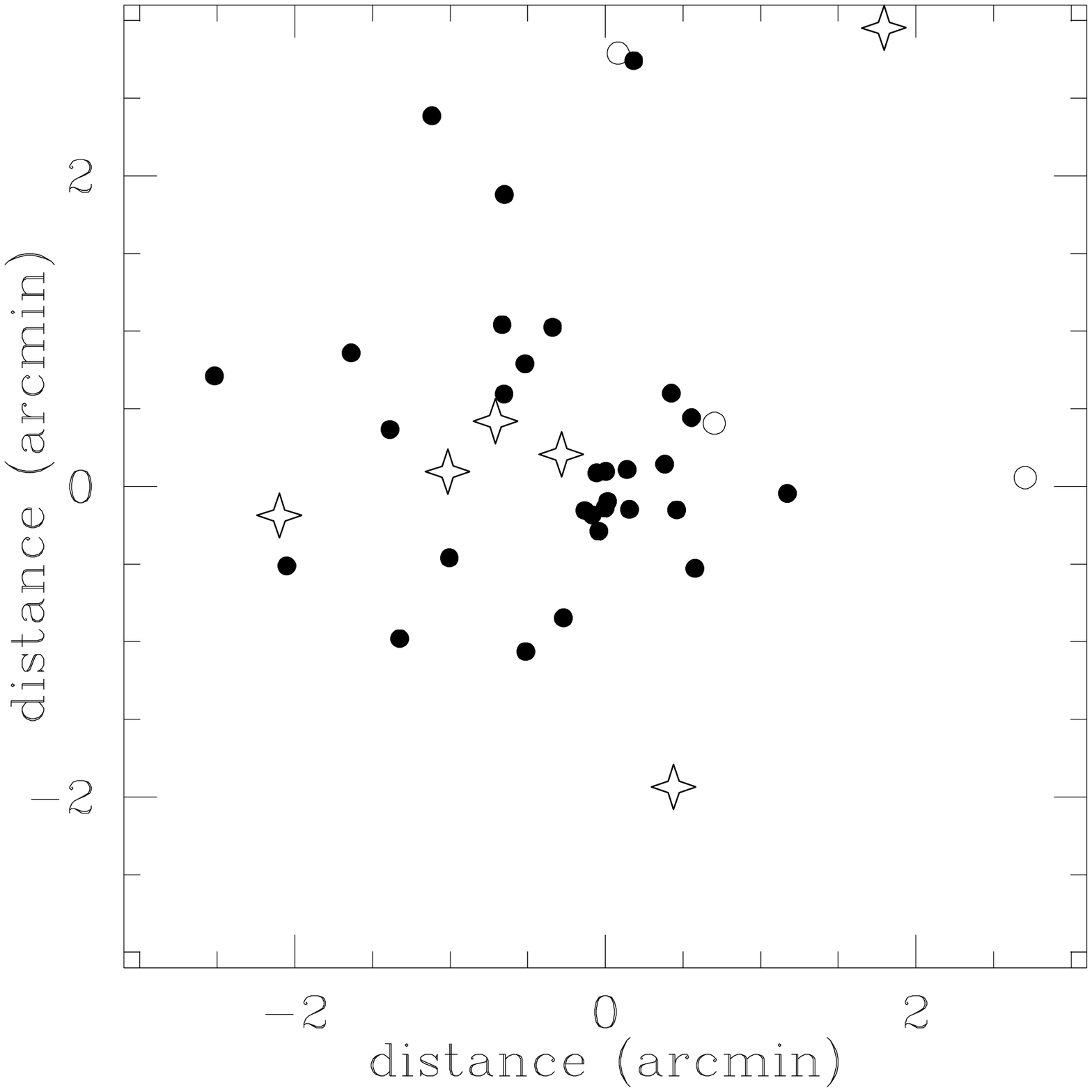,width=0.25\textwidth}
\psfig{file=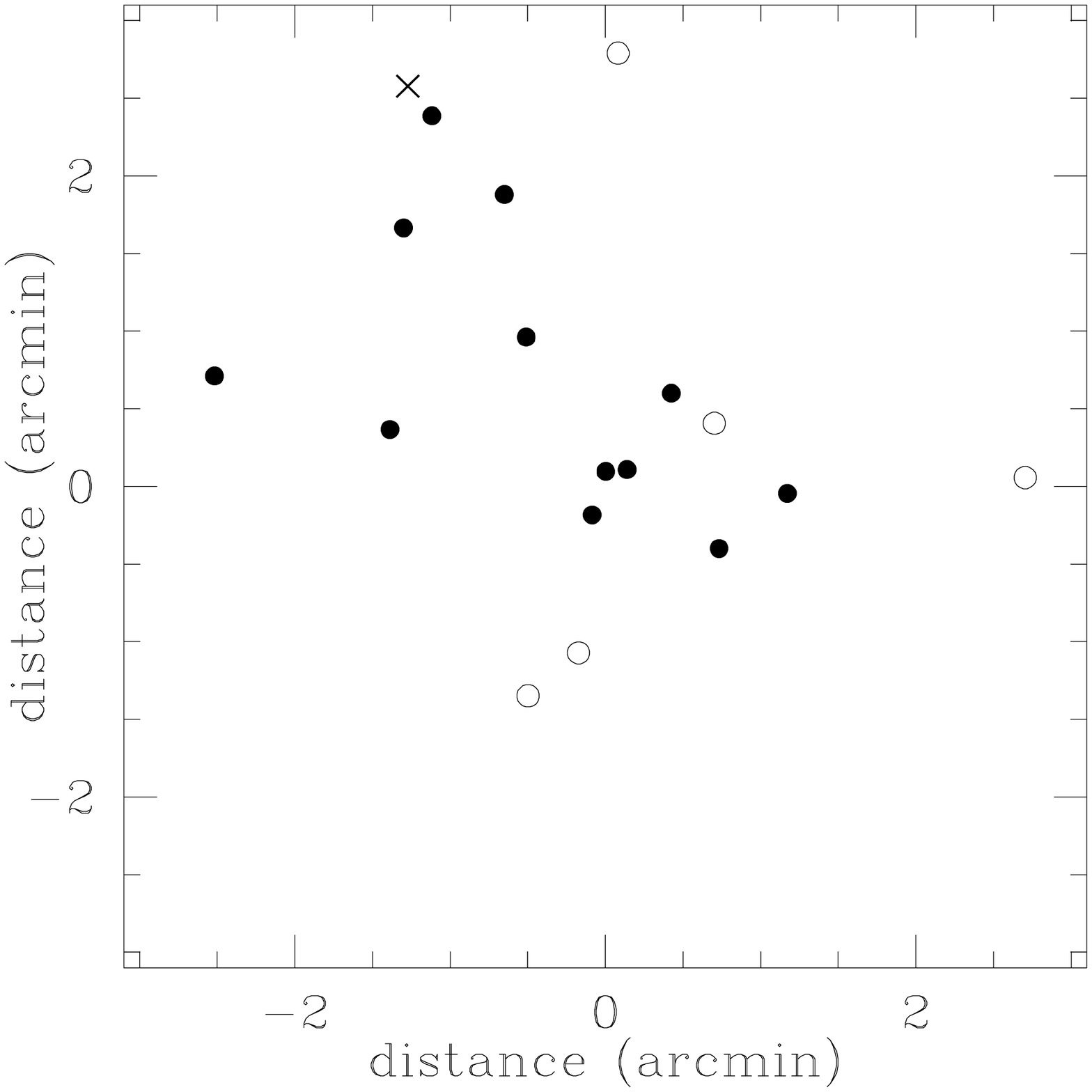,width=0.25\textwidth}}
}
\caption{
   On the top: fluxes of confirmed member galaxies detected with
  the two filters as a function of distance to the cluster center.  On
  the center: LW3/LW2 color versus the distance to the cluster center
  for LW2 (left) and LW3 (right) detections including upper limits.
  Fluxes and colors are binned in three bins and bootstrap technique
  has been used to evaluate 1$\sigma$ error-bands. Points mark fluxes
  or LW3/LW2 ratios for individual galaxies. Colors of template
  galaxies in Figure 8 are reported for reference (LW3/LW2 = 0.4, 1.0,
  2.0 and 5 for A, B, C and D template, respectively).  On the bottom:
  the spatial distribution of LW2 (left) and LW3 (right) detections.
  Confirmed cluster members are filled circles, background galaxies
  are empty circles, starlike symbols are stars and finally crosses
  are galaxies with unknown redshift.}

\label{ima:flux_vs_dist}
\end{figure}

For this reason we explored the relationship between the fluxes of
ISOCAM galaxies in the two bands and their distance from the cluster
center within the radius of $\sim 0.5$ Mpc, which roughly corresponds
to 2 core radii.  Figure~\ref{ima:flux_vs_dist} reports, for the
cluster members detected in the two bands, fluxes and colors as a
function of the radius and the spatial distributions of the
detections. Median values and 1$\sigma$ error bands have been computed
for fluxes and colors divided in three bins. Dispersions have been
computed using the bootstrap technique and taking into account also
upper limits in the case of color diagrams.  The figure shows that the
majority of bright LW2 sources are concentrated in the central region:
75\% of the sources within 1.3 arcminutes are brighter than 0.2 mJy,
while 70\% of the sources in the external region are fainter than 0.2
mJy. If we take into account that the sensitivity of the observations
decreases far from the cluster center, the real effect is even
stronger than our measurements indicate.  We can understand this fact
because LW2 detections correspond to the brightest optical galaxies,
since LW2 emission is dominated by stellar emission, and reflects the
distribution of galaxies in the cluster.

On the contrary, the emission in the LW3 band seems to increase with
distance from the cluster center. In the case of our confirmed members,
we have a significant correlation between LW3 flux and distance to the center(
the classical Spearman correlation coefficient, $\rho_S=0.52$, has the high 
92\% significance). 
We have also explored the color LW3/LW2 of sources detected in the
LW2 and LW3 band, using upper limits when sources are not detected in one
of the bands. We see that the mean color LW3/LW2 of LW3 detected galaxies increases with
the distance to the cluster center while that  of LW2 galaxies is almost constant.  
This behavior is consistent with the spatial distribution of
starburst galaxies and blue galaxies responsible for the
Butcher-Oemler effect. Therefore, we can conclude that selecting galaxies which
emit in the LW3 band allows one to emphasise star formation galaxies.

\subsection{Color Distribution}
\begin{figure}[t!]
\psfig{file=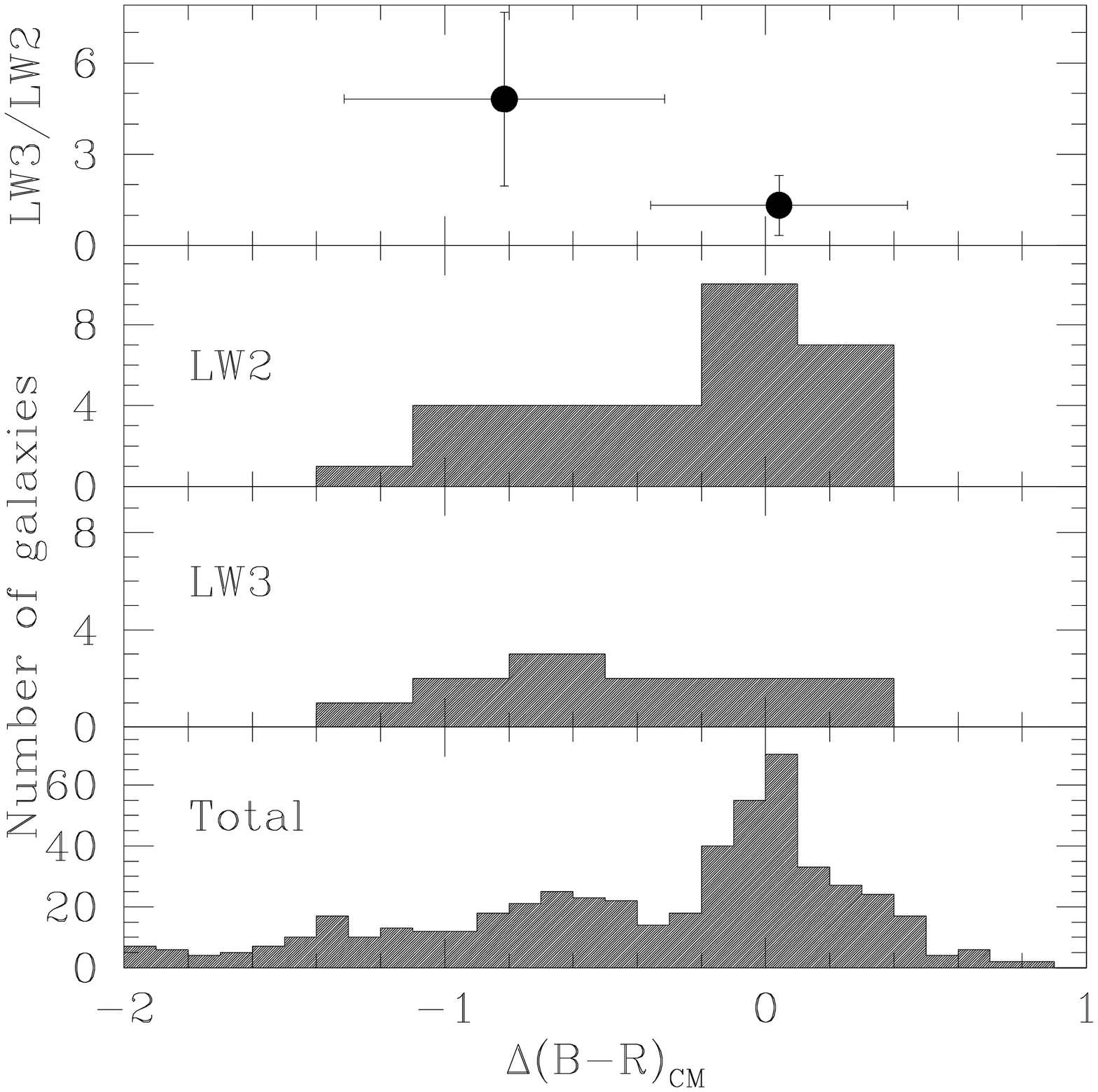,width=0.5\textwidth,angle=0}
\caption{On the bottom:
  distributions of differences between the B-R color of galaxies and
  the B-R color expected from the B magnitude if the galaxy would
  obey the color-magnitude relation. The LW2 distribution does not
  differ from the global galaxy distribution, while LW3 distribution
  does not show a peak corresponding to the color-magnitude relation.
On the top: the mean color of sources emitting in the two filters versus the
distance from color-mag relation in the B-R plane. Outliers of 
the color-mag relation have a LW3/LW2 mean color bigger than sources in the
color-mag relation. }
\label{ima:colors}
\end{figure}
Do the galaxies which emit in the MIR have peculiar optical
signatures ?  To answer this question we have studied the optical
colors of these galaxies with respect to the colors of the other
cluster galaxies (see figure~\ref{ima:colors}).  Colors and magnitudes
of early-type galaxies in galaxy clusters are strongly correlated
(Visvanathan \& Sandage, 1977).  Galaxies which do not obey this
correlation are believed to be late type or peculiar. In particular,
the blue outliers of this relationship are
responsible for the Butcher-Oemler effect.  To enhance the position of
ISOCAM galaxies in the color-magnitude plot, we have fitted the
color-magnitude relation of A1689 by using
a k-$\sigma$ clipping technique (see Duc \etal, 2000) 
and then we have computed, for each galaxy, the difference between its
B-R color and the color given by the color-magnitude relation at the
B magnitude of the galaxy.  Figure~\ref{ima:colors} shows the
distribution of these relative colors for the detected LW2 and LW3 galaxies
and for all the optical galaxies in the region observed by ISOCAM (total).
The figure showing the total distribution 
has a peak at zero which corresponds to the galaxies obeying
the color-magnitude relation.

While the distribution of LW2 galaxies is similar to the total one,
the LW3 distribution clearly misses the peak due to the
color-magnitude relation and most of the LW3 detections are
blue-outliers of the relation which are supposed to be responsible for
the Butcher-Oemler effect.

 Finally, if we consider the galaxies detected in
the two filters (see the top of Figure~\ref{ima:colors}), the mean LW3/LW2 color
of the galaxies obeying the color-magnitude relation is smaller than that
of the blue-outliers.

We can easily explain the absence of LW3 detections in the peak
because early-type galaxies are rarely detected with the LW3 filter
(see figure~\ref{ima:sed}) except for very bright galaxies in the
center of the cluster where the observations are highly sensitive.  On
the other hand, the blue color  and the high LW3/LW2 flux ratio of most  of
the LW3 detected galaxies imply that some of the blue galaxies
responsible for the Butcher-Oemler effect harbor hidden star
formation, as revealed by dust emission.  The color of these galaxies is still bluer than the
color-magnitude relation because only part of the young stars is
hidden by dust (see e.g. Poggianti et al., 1999).

\begin{figure}[t!]
\hbox{
\psfig{file=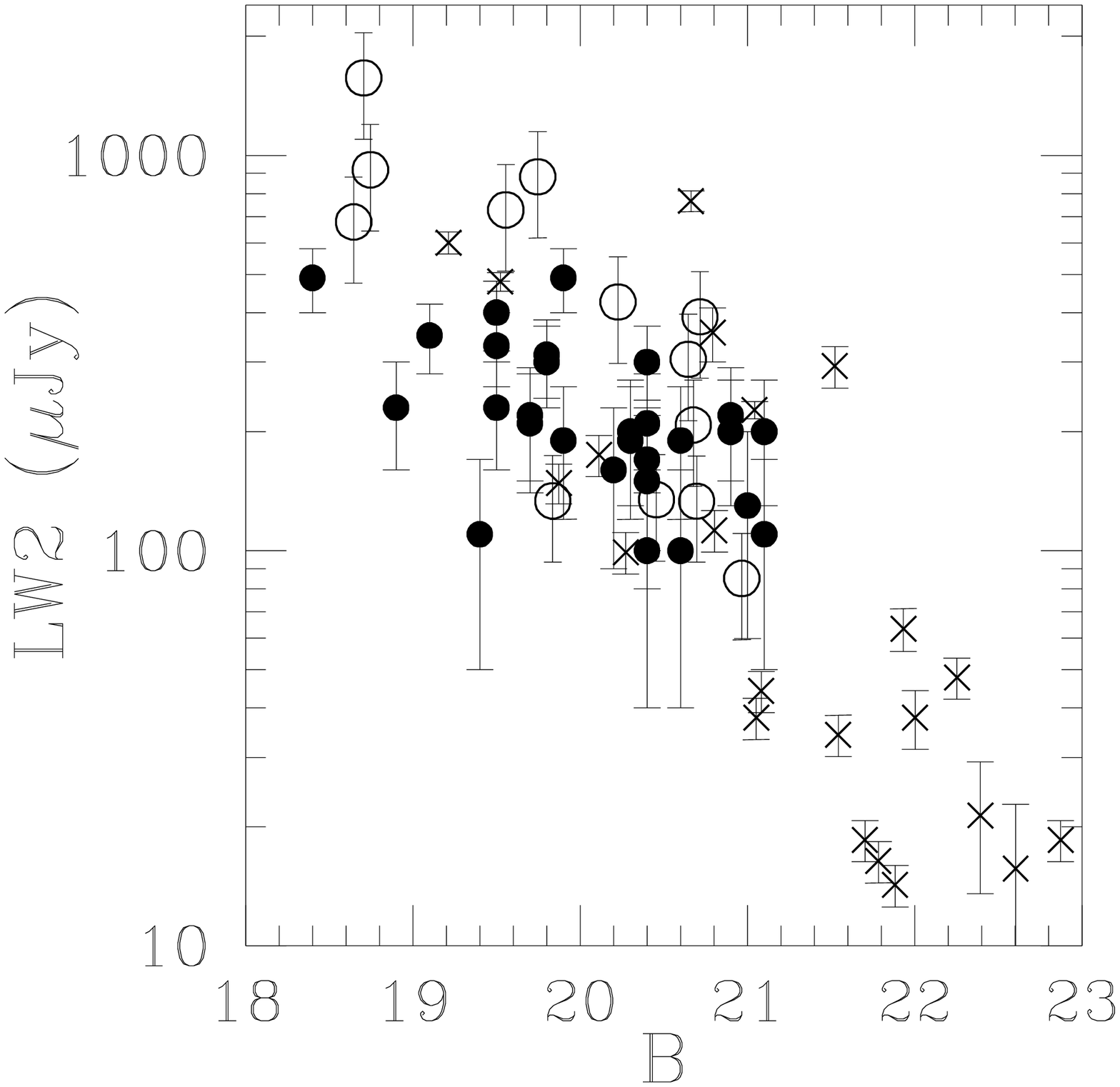,width=0.25\textwidth}
\psfig{file=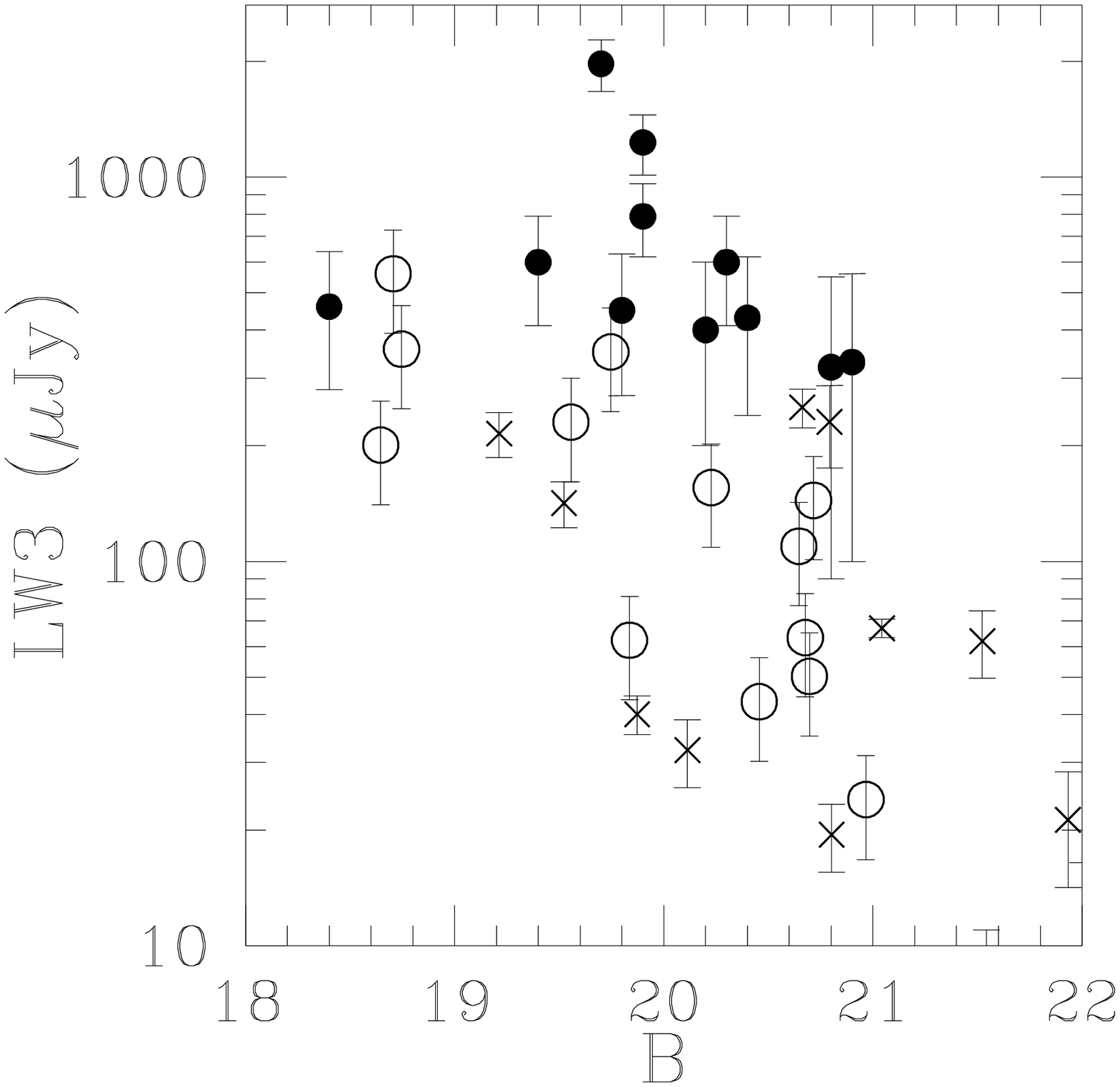,width=0.25\textwidth}
}
\caption{
   Comparison of fluxes of A1689 galaxies (filled circles) with
  those of Coma galaxies (empty circles) and Virgo galaxies (crosses)
  placed at the distance of A1689, as a function of their B magnitude.
  We note that, in the LW3 band, A1689 galaxies are on average
  brighter than local cluster galaxies.  }
\label{ima:flux}
\end{figure} 

\subsection{Comparison with local clusters and field galaxies}

Galaxies detected in the LW2 band correspond to the brightest optical sources,
which results in  an overdensity in the cluster region with respect to the
field.  

On the other hand, the density of galaxies detected in the LW3 band in the A1689 region is
clearly in excess with respect to the field.  By considering the
100\% completeness limit of 0.6 mJy, the expected number of sources
brighter than this limit in an area of 20 square arcminutes is
3$^{+2}_{-1}$, according to the integral counts of Elbaz \etal (1999),
while we find 9 galaxies above this flux, 5 of them are cluster
members. If we consider fluxes greater than 0.4 mJy (the 90\%
completeness limit), we expect $7^{+3}_{-2}$ background galaxies while
we find 14 galaxies, 9 of them are cluster members.

We compare in this paragraph the MIR fluxes and MIR/optical flux ratios of the 
A1689 galaxies with the galaxies detected in local clusters and in the
field.
 Our goal is to
detect an eventual excess of bright LW3 sources in A1689 with
respect to local rich clusters.  In fact, according to previous IRAS
studies of nearby clusters (see e.g. Bicay \& Giovanelli, 1987; Wang
et al., 1991), almost only ``IR-normal'' galaxies have been detected
in clusters in contrast to the field where at least 20\% have FIR
luminosities $> 10^{11}$~L${_{\odot}}$.  Unfortunately, only a few
such clusters have been observed in the MIR with ISOCAM.

Boselli \etal (1997) have studied a complete sample of late-type and
S0 galaxies brighter than $B_T \le 18$ in the Virgo cluster. 
For our comparison we use the subsample of 27 galaxies observed in a circle
of 2 degrees centered on M87, which corresponds to $\sim$ 0.5 Mpc, almost the 
same as in our ISOCAM observations. For most of the galaxies in this
region we used a new estimate of the fluxes as obtained by Roussel \etal
(2000).  

\begin{figure}[t!]
\hbox{
\psfig{file=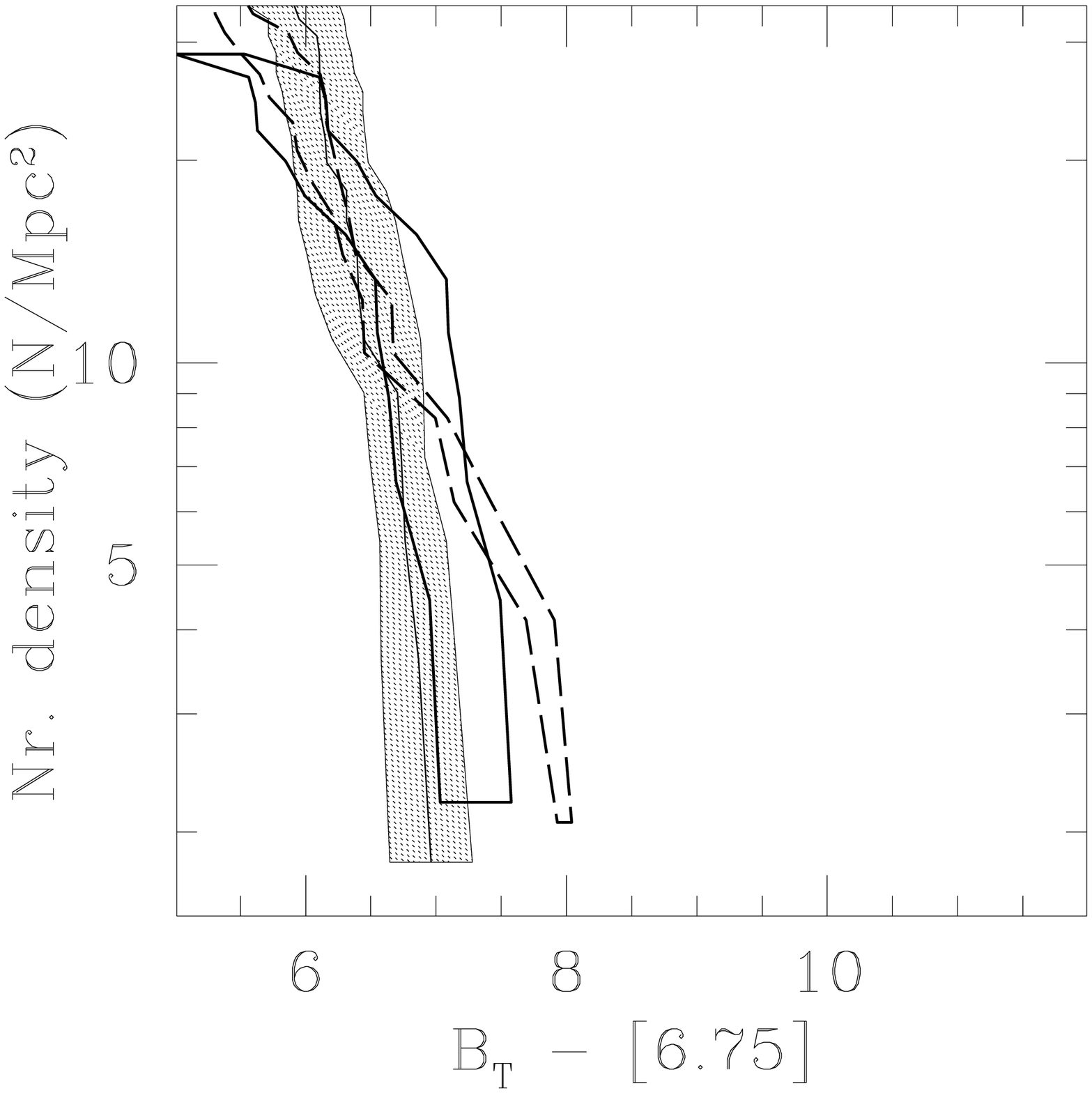,width=0.25\textwidth}
\psfig{file=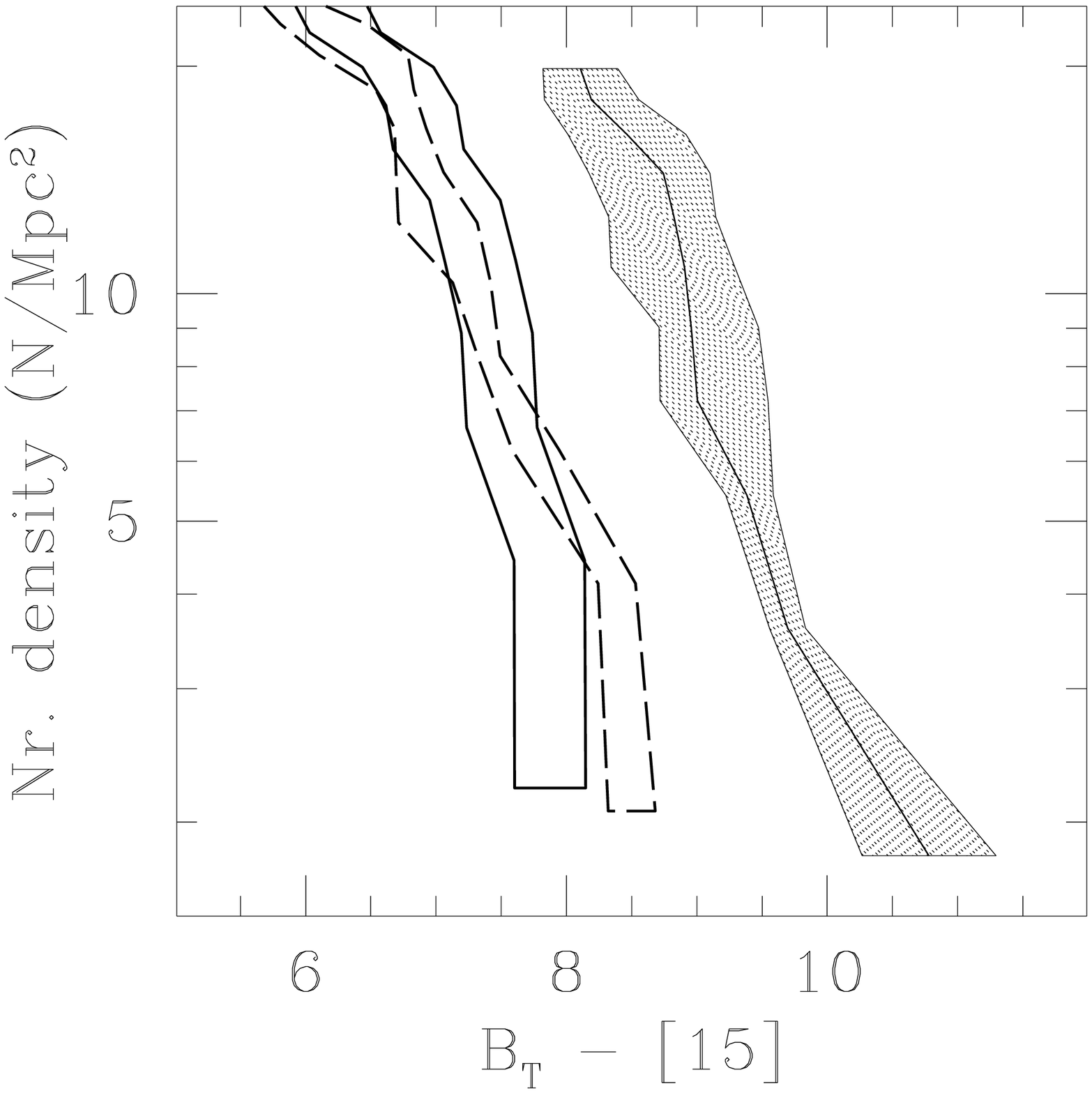,width=0.25\textwidth}
}
\caption{
Comparison among the cumulative distributions of the MIR/optical fluxes for
  galaxies in the central region ($R<0.5$ Mpc) of Virgo (dashed line),
  Coma (continuous line) and A1689 (hatched region) clusters. Coma and
  Virgo distributions are scaled at the richness of A1689 using their
  velocity dispersions as richness parameter. The $B_T -[6.75]$ color distributions
  (left) of the three clusters do not differ significatively. The
  $B_T-[15]$ color distribution (right) of A1689 shows an excess of galaxies with
  high $B_T-[15]$ colors with respect to the other two local clusters,
  which have a similar behavior.  }
\label{ima:opt_mir}
\end{figure} 

A sample of 13 spiral and irregular galaxies within 1 degree ($\sim$
1.5 Mpc) of the Coma cluster center has been studied by Contursi (1999).
The galaxies of the sample have been selected among the galaxies 
detected by IRAS at 60$\mu$m with blue colors ($B-H \le 2.75$).
The sample, although not complete, can be
taken as representative of the more active  galaxies in the
cluster.
Moreover, a sample of seven E+A galaxies has been observed in Coma
by Quillen \etal (1999). 
These galaxies appear to be fainter than the
late-type galaxies observed by Contursi (1999), which is in agreement
with their small contribution to the total FIR emission estimated by
Quillen \etal (1999).

Figure~\ref{ima:flux} shows the comparison of the MIR fluxes of
  A1689 galaxies with those of Virgo and Coma galaxies, placed at the
  distance of A1689 by computing the K-correction on the basis of the
  MIR SED of the photo-dissociation region NGC 7023 (Cesarsky \etal,
  1996b; see Fig.~\ref{ima:sed}). While the distribution of LW2 fluxes
  is similar for the three clusters, in the LW3 band A1689 galaxies
  appear brighter than galaxies in the local clusters.  The use of the
  SED of a starburst galaxy would increase this effect.

\begin{figure}[t!]
\hbox{
\psfig{file=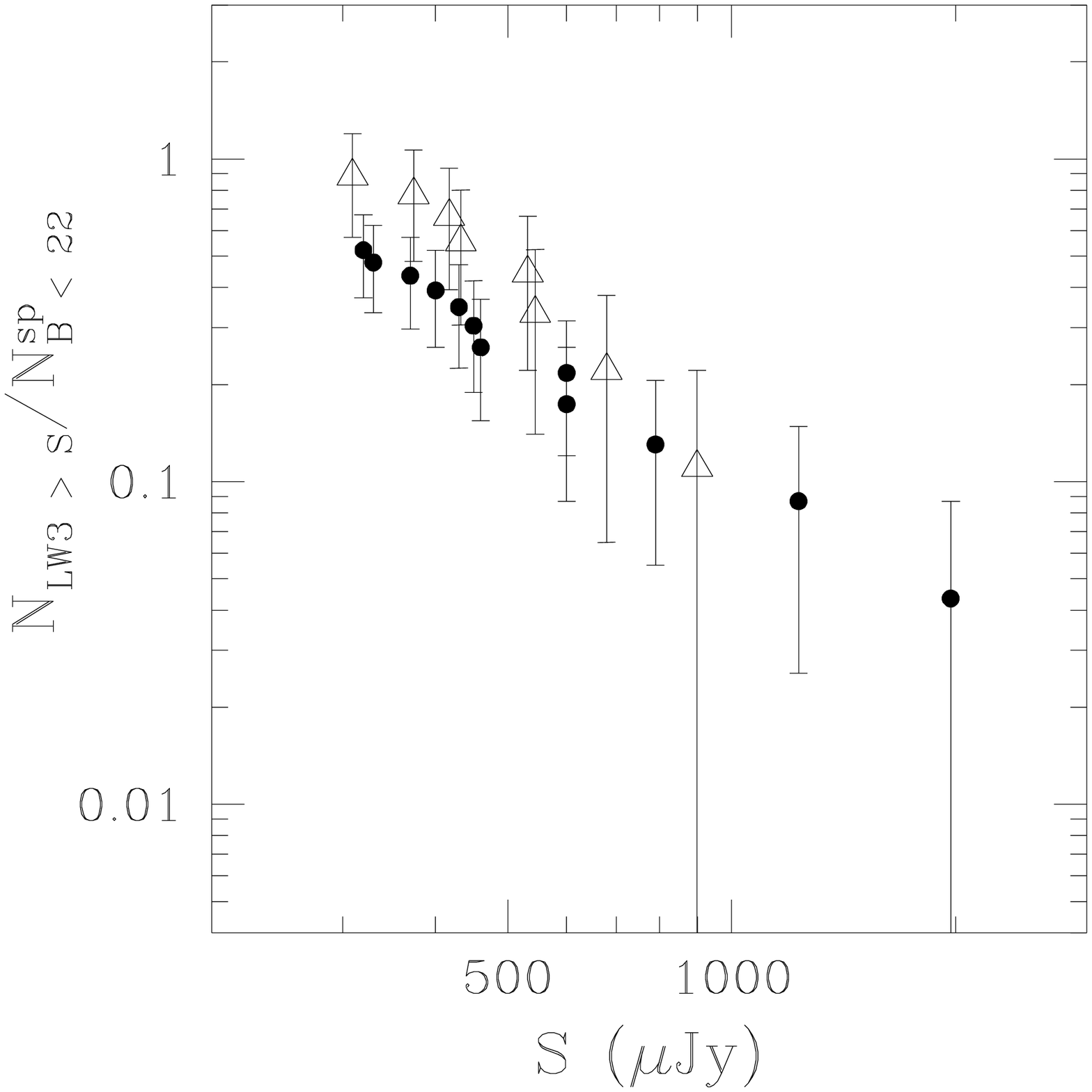,width=0.25\textwidth}
\psfig{file=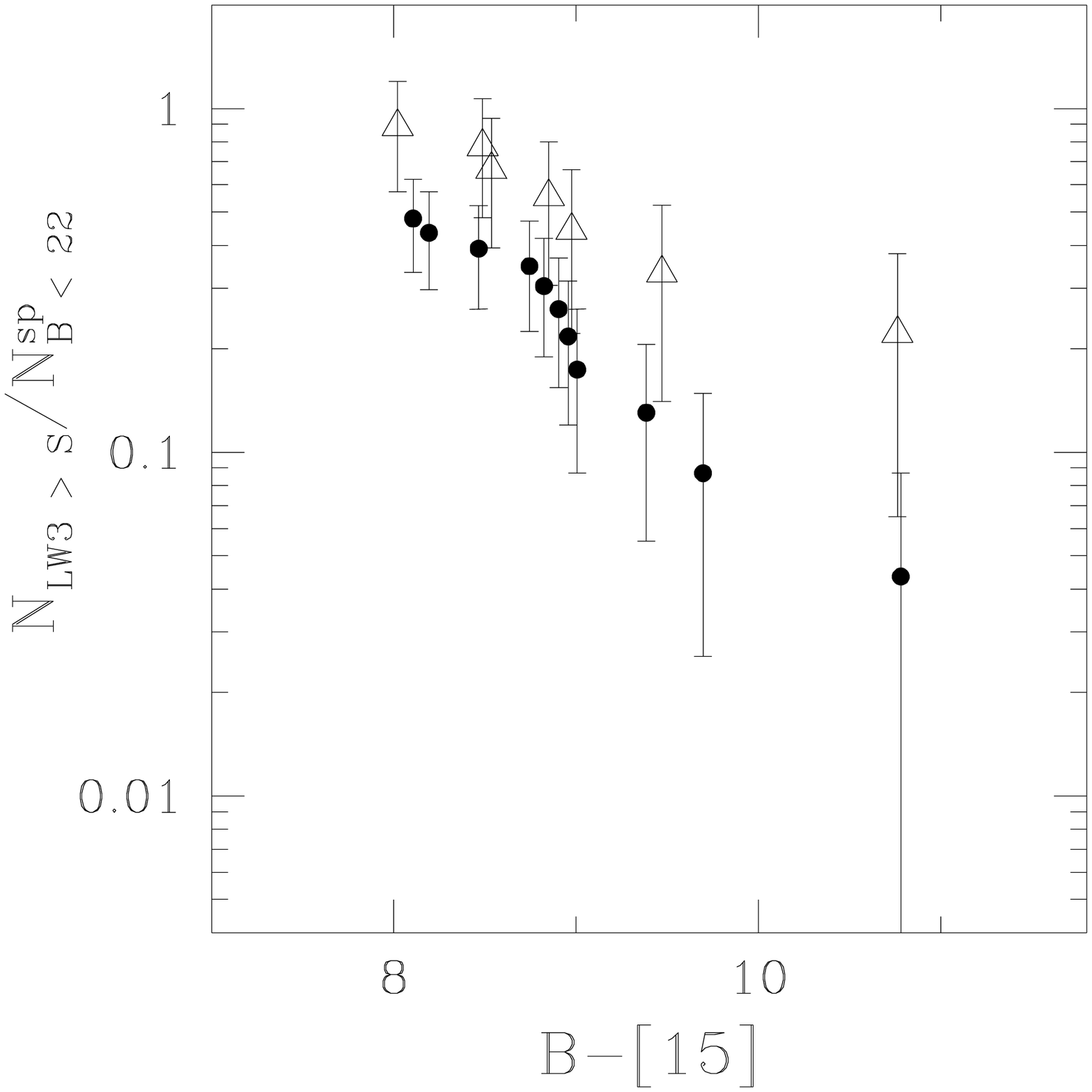,width=0.25\textwidth}
}
\caption{
  Comparison between LW3 fluxes (left) and LW3/optical colors (right) 
of A1689 (full circles) and CFRS galaxies (empty triangles) within the redshift range 0.1--0.3.
The cumulative distributions are normalized to the number of spirals with B$<$22 in the observed areas.
Poissonian error bars are shown.
  }
\label{ima:cfrs}
\end{figure}

A direct comparison of luminosity functions of the three cluster samples,
to deduce if there is a real excess of luminous 15$\mu$m
sources in A1689 with respect to local clusters, is difficult. In particular, it is difficult to correct
luminosity functions for incompleteness, due to the optical selection of galaxies
in the local clusters. Moreover, this comparison is affected also by the small size of the samples and
the  different cluster richness.

To overcome these problems, we have explored the ratio of MIR to
optical fluxes of the galaxies in the central region of the clusters
(radius of 0.5 Mpc), which is relatively free of sample selection.
Moreover, since in this region, which corresponds to the A1689 field
observed by ISOCAM, the bright end of Coma and Virgo MIR luminosity
functions are well sampled, we can scale the cumulative distributions
with respect to the richness of the clusters.  In this way it is
possible to also detect an excess in the density of luminous sources.
To do so, we scale the distributions at the richness of A1689 using
the velocity dispersions of the clusters, by exploiting the
linear relationship between the velocity dispersion and the richness of galaxy clusters 
within 0.5 Mpc (Bahcall et al. 1981).
 We use the values of $\sigma_r =
632^{+41}_{-29}$, $821^{+49}_{-38}$, $1429^{+145}_{-96}$ for Virgo,
Coma and A1689 velocity dispersions, respectively (Fadda et al. 1996,
Girardi et al. 1997).

We have transformed MIR fluxes in magnitudes using the relationships of
Omont et al. (1999):
\[\mbox{mag}(LW2) = [6.75] = 12.39 - 2.5 \log [F_{LW2} (mJy)]    \]
\[\mbox{mag}(LW3) = [15] = 10.74 - 2.5 \log [F_{LW3} (mJy)]    \]
 In order to put the galaxies of Coma and Virgo at
the distance of A1689, we have applied a K-correction to the 
B fluxes of Virgo and Coma galaxies according to Poggianti (1997).

Figure~\ref{ima:opt_mir} shows that the distributions of $B_T-[6.75]$ colors are similar 
for the three clusters, while the $B_T-[15]$ color distribution of A1689 differs significantly from
that of the other two clusters. A1689 shows an excess of galaxies with steep spectra
with respect to Coma and Virgo clusters. If we remember that LW3 sources
correspond to blue outliers of the color-magnitude relation of A1689 galaxies,
we can conclude that the LW3 emission is a very sensitive indicator of star formation.
A kind of analog of the optical Butcher-Oemler effect is clearly revealed by the ISO observations.

Finally, we have compared the galaxies of A1689 detected in the LW3 band
with field galaxies located at a similar redshift (figure~\ref{ima:cfrs} ).  In this
perspective, we used a sample of galaxies observed in the Canada
France Redshift Survey field CFRS-0300 (Flores \etal 2000) located at
redshifts between 0.1 and 0.3 in order to obtain sufficient
statistics. 

Since most of the sources which emit in the LW3 band at this redshift
are spiral or irregular galaxies, we have normalized the distributions
to the number of spiral/irregular galaxies in the field and in the
cluster with $B<22$.  In this way, we take out the effect of the
morphology-density relation.  For the field galaxies, we used the
percentage of spiral/irregular galaxies in the CFRS fields according
to Brinchmann et al. (1998).  For A1689, we used HST morphologies
(Couch, private communication) and the spectroscopic redshifts by Dye
et al. (2000) in the field observed by ISOCAM.
  
The two distributions are compatible within the error bars.  This
means that the 15$\mu$m luminosities and the B-[15] colors of A1689
galaxies detected in the LW3 band are similar to those of field
galaxies at the same redshift.

15$\mu$m sources are typically spiral galaxies which constitute
  20\% of the total population of A1689 galaxies and 50\% of field
  galaxies at the redshift of the cluster. On the other hand, by
  comparing counts in the V band of field galaxies and galaxies in the
  central region of A1689, Wilson et al. (1997) find an overdensity of
a factor  two for $V<22$. 
Since we find a clear overdensity of 15$\mu$m sources in the cluster with
respect to the field, we can say that the environment of A1689 likely triggers
starburst episodes in galaxies in the cluster outskirst, although the overall
properties (infrared luminosities and FIR/optical colors) of these are similar to field starbust galaxies.

\begin{table}[t!]
\begin{tabular}{cccc}
\hline
Solar Elongation & Filter & Zod. light (model) & Background\\
(deg) &  & (MJy/sr) & (MJy/sr)\\
\hline
71.5 & LW2 & 9.5 & 9.7 $\pm$ 0.2\\
71.5 & LW3 & 64.6 & 54.5 $\pm$ 0.6\\
104.6 & LW3 & 39.5 & 34.5 $\pm$ 0.4 \\
108.6 & LW2 & 4.3 & 4.7 $\pm$ 0.1\\
72.9 & C200 & 3.8 & 4.0$\pm$0.2\\
\hline
\end{tabular}
\caption{
Measured background values compared with zodiacal light previsions
using DIRBE measurements. Almost all the background emission can be explained
with zodiacal light.}
\label{tab:dirbe}
\end{table}

\subsection{Diffuse IR emission}

In this section we address the issue of the possible presence of
diffuse infrared emission due to intra-cluster dust.  Theoretically it
is difficult to understand how such intra-cluster infrared emission
could originate from the central region of rich and well evolved
galaxy clusters since the timescale for the destruction of dust grains
shocked by the hot X-ray plasma of electrons (sputtering) is of the
order of a few $10^8$ years (Dwek, Rephaeli \& Mather 1990).  Only
dust replenishment by galactic winds or by stripping of
  in-falling galaxies could explain such emission but it is not
expected to happen in the central Mpc of rich clusters.  However,
a systematic study of 56 clusters of galaxies using IRAS data led Wise
\etal (1993) to suggest detections for intracluster emission in two
clusters. More recently Stickel \etal (1998) claimed to have detected
intracluster dust emission in the central region of Coma using ISOPHOT
at 120\mic.

Since A1689 and Coma have similar properties, one may expect similar
diffuse emission in the two clusters.  We have measured the diffuse
emission in the field of A1689 in three bands: LW2 and LW3 with ISOCAM
and at 200\mic with ISOPHOT.  In Table~\ref{tab:dirbe}, we compare
our values with the expectations based on a model of the zodiacal
background light using the DIRBE observations (B. Reach, private
communication).  All three values agree well with the DIRBE estimates
for this region.

Unfortunately A1689 is very close to the ecliptic plane (at an
ecliptic latitude of $\beta =$ 5.78 degrees) where the infrared
background due to diffracted solar light is very high. Hence if a
small intra-cluster contribution were present it would be very
difficult to detect. In particular, a FIR background light of 0.1
MJy/sr as proposed by Stickel \etal (1998) for Coma is still
compatible with our error bars. However, we can use the MIR flux
densities of the A1689 galaxies resolved with ISOCAM to deduce their
contribution to the FIR background light assuming a given FIR over MIR
ratio. We report this contribution in Table~\ref{tab:FIR} using the
ratio reported by Contursi (1999) for the Coma galaxies for which
there is both ISOCAM MIR and IRAS FIR measurements. We find values of
the order of the measurement of Stickel \etal (1998), in agreement
with Quillen \etal (1999) for Coma. Hence, our results agree with the
conclusion of Quillen \etal (1999): the origin of the FIR background
light in the direction of rich clusters is most probably due to the
integrated contribution of the cluster galaxies.

\begin{table}[t!]
\begin{tabular}{cccc}
\hline
PHOT & FIR (LW2 $>$ 0.2mJy) & FIR (LW3 $>$ 0.5mJy) \\
band &  (MJy/sr)&(MJy/sr)\\
\hline
120\mic & 0.22 & 0.24\\
175\mic & 0.14 & 0.16\\
200\mic & 0.09 & 0.11\\
\hline
\end{tabular}
\caption{
Estimates of the emission in several PHOT bands in the FIR 
by galaxies brighter than 0.2 mJy in the LW2 band
and than 0.5 mJy in the  LW3 band.
}
\label{tab:FIR}
\end{table}
\section{Summary and Conclusions}
We have presented the results of  infrared observations of
Abell 1689, an exceptionally rich cluster of galaxies at intermediate
redshift ($z\simeq 0.181$), with ISOCAM, at 6.7 and 15\,$\mu$m, and
ISOPHOT at 200\,$\mu$m, from the {\it Infrared Space Observatory}
(ISO). 

The ISOCAM observations, which cover a region of 20 square arcminutes,
reach a sensitivity limit of 0.15 mJy at 6.7\mic and 0.3 mJy at 15\mic.
From a spectroscopic follow-up (Duc \etal 2000) and a set of photometric
redshifts (Dye \etal 2000) we know redshifts for all the galaxies except one.
This has allowed us to 
explore the infrared properties of cluster members within a radius of $\sim 0.5$ Mpc.

Galaxies  detected in the LW2 band, which is mostly dominated by
stellar emission, correspond to the brightest optical objects. They
appear concentrated in the central part of the cluster and show
optical colors similar to those of the overall cluster population.

The cluster members detected in the LW3 band  are blue
outliers of the cluster color-magnitude relation and become brighter 
going from the center to the outer parts of the cluster.

 We have estimated the cumulative distributions of the ratios of
  MIR to optical fluxes in the two bands above the 90\% completeness
limits of 0.2 mJy for LW2 and 0.4 mJy for LW3.  Comparing these
  distributions to those of the two nearby galaxy clusters, Virgo and
  Coma, scaled at the richness of A1689, we find that they are
  compatible in the LW2 filter.  On the contrary, Abell 1689 shows a
  systematic excess of galaxies detected at 15\mic with high 15\mic to
  optical ratio indicating higher star formation activity.
Therefore, the far-infrared shows a trend similar to the
Butcher-Oemler effect measured in the optical. The case of A1689 is
even more remarkable since it shows an excess of activity in the
cluster galaxy population at $z\sim0.2$ with respect to nearby
clusters which is still debated on the basis of optical data (Gudehus
\& Hegyi 1991).  A comparison with field galaxies observed in the LW3
band at the redshift of A1689 shows that, in the limits of our fairly
large error bars, cluster galaxies are not more luminous and more
active than field galaxies.

 Moreover, we find a clear overdensity of 15$\mu$m sources in the
cluster with respect to the field while optical counts of spiral
galaxies which mainly emit at 15$\mu$m do not show this overdensity.
Therefore, we can say that the environment of A1689 likely triggers
starburst episodes in galaxies in the cluster outskirst, although the
overall properties (infrared luminosities and far-infrared/optical
colors) of these are similar to field starbust galaxies.

The spatial resolution of ISOPHOT is too poor to study the galaxy
population. We used it to calculate the integrated FIR
background light of the cluster and compared it to the expected contribution of
the individual galaxies, evaluated using their MIR luminosities and
assuming a FIR/MIR ratio based on Coma galaxies. We find no
evidence for intra-cluster dust emission in the limit of our error
bars, whereas the integrated FIR light produced by the individual
galaxies seems sufficient to produce an integrated FIR flux at a level similar 
to that previously reported for the Coma cluster.

\begin{acknowledgements}
  We thank Ren\'e Gastaud and Herv\'e Aussel for help with the data
  reduction of the ISOCAM images, Guilain Lagache and Marc Sauvage for
  their advice on ISOPHOT data reduction and Bianca Maria Poggianti
  and Suzanne Madden for interesting discussions and suggestions when
  we were writing the manuscript.  We are also grateful to Bill Reach
  for relevant values of zodiacal light produced by his model, to
  Warrick J. Couch who provided us the HST morphologies of A1689
  galaxies prior to publication and to Simon Dye who gave us their
  photometric redshifts.  We thank the referee Ian Smail for his
  careful reading of the manuscript and his helpful suggestions.
  P.--A.  D.  acknowledges support from the network Formation and
  Evolution of Galaxies set up by he European Commission under
  contract ERB FMRX--CT96086 of its TMR program.
\end{acknowledgements}

\end{document}